\documentclass[twocolumn]{aastex631}

\usepackage{natbib}
\usepackage{mathtools}
\usepackage{amsmath,bm}
\usepackage{hyperref}
\usepackage{color}
\usepackage{multirow}
\usepackage{gensymb}
\usepackage{lipsum}

\DeclareMathAlphabet{\pazocal}{OMS}{zplm}{m}{n}
\newcommand{\unif}{\pazocal{U}}
\DeclareMathOperator{\atantwo}{atan2}
\newcommand{\round}[1]{\ensuremath{\lfloor#1\rceil}}

\shorttitle{Full-sky model of Galactic dust based on filaments}
\shortauthors{Herv\'ias-Caimapo \& Huffenberger}

\begin{document}

\correspondingauthor{Carlos Herv\'ias-Caimapo}
\email{cherviascaimapo@fsu.edu}

\title{Full-sky, arcminute-scale, 3D models of Galactic microwave foreground dust emission based on filaments}

\author[0000-0002-4765-3426]{Carlos Herv\'ias-Caimapo}
\author[0000-0001-7109-0099]{Kevin M. Huffenberger}
\affiliation{Department of Physics, Florida State University, Tallahassee, Florida 32306, USA}

\begin{abstract}


We present the \textsc{DustFilaments} code, a full-sky model for the millimeter Galactic emission of thermal dust.  Our model, composed of millions of filaments that are imperfectly aligned with the magnetic field, is able to reproduce the main features of the dust angular power spectra at 353\,GHz as measured by the Planck mission. Our model is made up of a population of filaments with sizes following a Pareto distribution $\propto L_a^{-2.445}$, with an axis ratio between short and long semiaxes $\epsilon \sim 0.16$ and an angle of magnetic field misalignment with a dispersion RMS($\theta_{\rm LH}$)$=10$\degree. 

On large scales, our model follows a Planck-based template.  On small scales, our model produces spectra that behave like power laws up to $\ell \sim 4000$ or smaller scales by considering even smaller filaments, limited only by computing power.  We can produce any number of Monte Carlo realizations of small-scale Galactic dust. Our model will allow tests of how the small-scale non-Gaussianity affects CMB weak lensing and the consequences for the measurement of primordial gravitational waves or relativistic light relic species.

Our model also can generate frequency decorrelation on the modified blackbody spectrum of dust and is freely adjustable to different levels of decorrelation. This can be used to test the performance of component separation methods and the impact of frequency spectrum residuals on primordial B-mode surveys.  The filament density we paint in the sky is also able to reproduce the general level of non-Gaussianities measured by Minkowski functionals in the Planck 353\,GHz channel map.


\end{abstract}

\section{Introduction} \label{sec:intro}


The presence of Galactic foregrounds at millimeter wavelengths is one of the main hurdles for cosmology with the cosmic microwave background (CMB). This is especially true for the potential detection of a background of gravitational waves from inflation that source a curl polarization component in the CMB, also known as B-modes \citep{1997PhRvL..78.2058K, 1997PhRvL..78.2054S, 2016ARA&A..54..227K}. A measurement of the tensor-to-scalar ratio $r$ could substantiate or rule out different models of inflation \citep{2009arXiv0907.5424B}. This primordial signal peaks at degree scales, and at these scales, we already have full-sky observations, like the ones by the Planck experiment. These are not sensitive enough to detect the B-modes, but let us begin to construct large-scale models of foreground emission \citep{2016A&A...594A..10P}.
In addition to the foregrounds, the primordial B-mode signal is contaminated by B-modes generated via gravitational lensing of the CMB photons by large-scale structure between us and the surface of last scattering \citep{2006PhR...429....1L}. The B-mode signal from foreground contamination and from  gravitational weak lensing are each larger than the possible primordial cosmological B-mode signal. Constraining this lensed B-mode signals with arcminute-scale CMB data is vital to remove the lensing contaminant at degree scales

The new generation of CMB ground-based experiments will observe with high resolution over a huge fraction of the sky with very good sensitivity. Experiments like Simons Observatory \citep{2019JCAP...02..056A} and CMB-S4 \citep{2016arXiv161002743A} aim to observe $f_{\rm sky} \geq 0.4$ of the sky with an $\sim 6$\,m aperture, which is equivalent to an $\sim 1\arcmin$ resolution at 150\,GHz. At these small scales, the Galactic foregrounds, such as thermal dust and synchrotron, will have non-Gaussian features.  Structure in the emission originates from nonlinear processes in the interstellar medium (ISM) by the interaction of turbulence, energy injection, and the Galactic magnetic field. Methods for component separation and lensing reconstruction could suffer from unexpected non-Gaussian foregrounds and leave residuals in the science products. These residuals could potentially damage our efforts, so our methods must be tested against models that include them.

In the last few years, several models of the diffuse and extragalactic foregrounds at millimeter frequencies have been developed to simulate observations for CMB end-to-end pipeline data analysis \citep[e.g.][]{2008MNRAS.388..247D, 2013A&A...553A..96D, 2016MNRAS.462.2063H, 2017MNRAS.469.2821T, 2017MNRAS.464.3486Z}. However, these models are usually based in foreground templates in intensity and polarization observed directly by experiments like Planck. Since observations of the millimeter sky have limited resolution, the small scales of these templates are usually filled by generating Gaussian anisotropies with  power spectra that follow an extrapolation of the measured foregrounds at the large scales. This will usually take the form of a power law. Obviously, the problem with this approach is that it will not simulate a deviation from Gaussianity in polarized foregrounds, which is most likely present in the real sky.

Several works have looked at analyzing and quantifying the non-Gaussianity and statistical isotropy violation in radio and millimeter diffuse Galactic foregrounds \citep[e.g.][]{2013JCAP...02..031C, 2014PhRvL.113s1303K, 2015JCAP...11..019B, 2016JCAP...09..034R, 2018MNRAS.481..970R, 2019JCAP...10..056C, 2019MNRAS.487.5814V, 2021arXiv210400419R,2021A&A...649L..18R,2021ApJ...910..122S}. In general, they find that their deviations from Gaussianity are increased toward the Galactic plane at $\lesssim$degree scales. However, the lack of adequate resolution and signal-to-noise ratio prevents us from making conclusive statements at the few arcminute scales.

In particular, diffuse thermal dust emission from our galaxy, the subject of this work, is radiation from dust grains in the ISM. 
The polarization of the thermal dust is the product of the interplay of elongated dust grains aligned with respect to the Galactic magnetic field \citep{2003ARA&A..41..241D}. Turbulent, supersonic flows in the ISM compress the gas and organize it into a weblike structure of filaments \citep{2014prpl.conf...27A}. These filaments have been observed in multiple frequencies by many experiments, in particular at millimeter emission by Planck \citep{2016A&A...586A.135P,2016A&A...586A.136P,2016A&A...586A.141P}. Filamentary structure is measured in the Galactic H I emission and correlates well with the thermal dust polarization in the Planck 353\,GHz emission \citep{2014ApJ...789...82C, 2015PhRvL.115x1302C}. 

The CMB community has recently started to focus on developing millimeter foreground models with non-Gaussian small-scale emission. For example, some efforts have been focused on magnetohydrodynamic (MHD) simulations. Several works have looked at the effect in the ISM of turbulence driven by different processes (supernova explosions, massive star outflows, etc.) and how they shape the physical parameters such as the magnetic field and density and examined the Alfv\'enic and sonic Mach numbers of the flow \citep[e.g.][Stalpes et al. in preparation]{2018PhRvL.121b1104K,2019ApJ...880..106K,2020ApJ...894L...2B}. 

Other models have tried less computationally intensive methods to account for the three-dimensional structures of the Galactic magnetic field, layers of Galactic dust, and the spiral structure of the Milky Way \citep[e.g.][]{2011A&A...526A.145F,2017A&A...603A..62V,2018A&A...614A.124L,2018MNRAS.476.1310M}. Recently, a new approach has taught neural networks to extrapolate foregrounds from large to intermediate scales, then used that same extrapolation to go from observed intermediate scales in Planck to unobserved arcminute scales \citep[e.g.][]{2021ApJ...911...42K,2021MNRAS.504.2603T}. Another approach is to construct models based on observations specifically exploiting ancillary data such as Galactic H I emission, which will add information on a third radial dimension along the line of sight (LOS) using a Doppler velocity shift of the molecular gas in the ISM \citep[e.g.][]{2017A&A...601A..71G,2019ApJ...887..136C,2020A&A...640A.100A}.

Another interesting phenomenon discovered recently is the frequency decorrelation of the dust spectral emission, meaning that the flux between two or more frequencies is not a simple multiplicative factor but varies across and along LOSs. 
The dust in the galaxy will have different physical conditions, such as dust grain population, gas cloud velocity, direction of magnetic field, etc., which will generate an overlap of different frequency spectra. Frequency decorrelation was first analyzed in Planck data by \citet{2016A&A...586A.133P} and measured by \citet{2017A&A...599A..51P}.
It has been discussed in \citet{2018PhRvD..97d3522S}, \citet{planck_2018_xi}, and \citet{2021A&A...647A..16P}. Planck has measured limits on the decorrelation between the 217 and 353\,GHz channels for a large fraction of the sky, while \citet{2021A&A...647A..16P} measured the decorrelation on individual LOSs located at the Galactic pole regions.

In this work, we build a foreground model from the idea presented in \citet{filament_paper} that filaments and their interaction with the magnetic field can explain most of the features measured by Planck in the dust power spectra in \citet{planck_2018_xi}. While \citet{filament_paper} considered an idealized population of filaments and integrated their distributions to predict their power spectra in a semianalytic computation, here we create simulated populations of individual randomized filaments and combine the emission of each filament to produce a full-sky image of the Galactic thermal dust intensity and polarization at millimeter frequencies.

We produce a model that can reproduce the dust angular power spectrum and its features as measured by Planck in \citet{planck_2018_xi}. In particular, we look at the power law fit of the TE, EE, and BB spectra, as well as the $\mathcal{D}_{\ell}^{\rm BB}/\mathcal{D}_{\ell}^{\rm EE}$, $\mathcal{D}_{\ell}^{\rm TE}/\mathcal{D}_{\ell}^{\rm EE}$, and $r_{\ell}^{\rm TE}=\mathcal{D}_{\ell}^{\rm TE}/\sqrt{\mathcal{D}_{\ell}^{\rm TT} \mathcal{D}_{\ell}^{\rm EE}}$ ratios. We also introduce a simple method for generating frequency decorrelation, measured by the correlation $\mathcal{R}_{\ell}^{\rm BB}(217,353)$.  We are also able to reproduce the general level of non-Gaussianity in intensity, which we measure using Minkowski functionals (MFs).

This paper is organized as follows. In Section~\ref{sec:data}, we briefly present the Planck data we use to inform our model. In Section~\ref{sec:method}, we describe the method that generates the simulated map of thermal dust composed of individual filaments (with extra details appearing in Appendix~\ref{sec:appendix-method}). In Section~\ref{sec:results}, we present our results and compare our filament model to the Planck observations in detail. In Section~\ref{sec:discussion}, we discuss specific details about where our models and the observed sky might not match. Finally, in Section~\ref{sec:conclusions}, we draw our conclusions.


\section{Data that inform our model} \label{sec:data}
We build our model with public data from Data Releases (DRs) 2 and 3 from the Planck mission.

The main results we aim to reproduce in this paper are the power spectrum properties of the thermal dust emission in \citet{planck_2018_xi}. We use their same inputs, namely, the 217 and 353\,GHz frequency maps from the High-Frequency Instrument \citep[HFI;][]{2020A&A...641A...3P}, and both the full mission maps and the two half-mission splits for both frequency channels.  We also use the Intensity and Polarization Large Region (LR) 71 masks \citep[the polarization mask is shown in Fig.~2 of][]{planck_2018_xi} to estimate the power spectra in the same sky fraction when comparing to our model. 
For the LR71 mask and at an anchor scale of $\ell=80$, their measured ratios are $\mathcal{D}_{\ell}^{\rm BB}/\mathcal{D}_{\ell}^{\rm EE}=0.53\pm0.01$ and $\mathcal{D}_{\ell}^{\rm TE}/\mathcal{D}_{\ell}^{\rm EE}=2.77\pm0.05$. The reference value for the $\rm TE$ correlation is $r_{\ell}^{\rm TE} \sim 0.357$ for the LR71 mask. The measured slopes are $\alpha_{\rm EE}=-2.42\pm0.02$, $\alpha_{\rm BB}=-2.54\pm0.02$ and $\alpha_{\rm TE}=-2.50 \pm 0.02$. All of these amplitudes and slopes are somewhat mask-dependent.

We also use the component-separated thermal dust products derived from the GNILC method in \citet{2016A&A...596A.109P}, constructed from the DR2 LFI and HFI Planck maps \citep{2016A&A...594A...6P,2016A&A...594A...8P}. These data products include maps of thermal dust temperature and emissivity index found by fitting a modified blackbody (MBB) to the high-frequency $\nu > 353$\,GHz Planck maps. Finally, we use the thermal dust $Q$ and $U$ maps produced with the same GNILC method as above but with the DR3 maps \citep{2020A&A...641A...4P}. We use the thermal dust template with a uniform resolution of $80\arcmin$.

We also need a model for the Planck-measured 353\,GHz $\mathcal{D}_{\ell}^{\rm TT}$ to aid in the modeling of the temperature-to-polarization correlation. \citet{planck_2018_xi} did not provide a fit to it. We compute it with the \textsc{namaster} code \citep{2019MNRAS.484.4127A}, which we use for all of the angular power spectra in this work. We calculate the cross-spectra between the two half-mission maps (each with independent noise realizations) and subtract the CMB contribution by removing the best-fit theory CMB spectra from the Planck DR2 \citep{2016A&A...594A..11P}. We fit a power law to the remaining spectrum using the LR71 mask and using the Knox formula \citep{1997ApJ...480...72K} to account for the bandpower error bars. This fit is performed in the multipole range $260 \leq \ell <600$, where the Planck dust TT spectrum looks like a stable power law (and avoiding an oscillation around $\ell \sim 150$). Our model is given by
\begin{equation}
    \mathcal{D}_{\ell}^{\rm TT} = A^{\rm TT} (\ell / 80)^{\alpha_{\rm TT} + 2} \text{,}
\end{equation}
where we find $A^{\rm TT} = 28,097 \pm 1215$\,$\mu$K$^2$ and $\alpha_{\rm TT} = -2.60 \pm 0.03$.



\section{Method} \label{sec:method}

To generate realizations of our model, we populate an observer-centered volume with simply defined filaments. We fill a $(400\,{\rm pc})^3$ cube with a magnetic field composed of two parts: a dominant, correlated, isotropic, random component and a sub dominant, large-scale component based on the Jansson \& Farrar model \citep{2012ApJ...757...14J,2012ApJ...761L..11J}. The filaments are coherently oriented using the magnetic field. Following \citet{filament_paper}, we set the properties of the filament population---including the distribution of filament sizes, aspect ratios, and polarization fractions---so that the resulting power spectra reproduce the observations by Planck.  We integrate the filament density profiles and magnetic field along the LOS to generate maps of the intensity and linear Stokes parameters.  From these maps, we verify that we have achieved the target power spectra and other properties. 

\begin{deluxetable*}{llll}
    \tablecaption{Parameters We Adopt for our Thermal Dust Model. 
    \label{table:parameters}
    }
    \tablehead{ \colhead{Parameter}  & \colhead{Symbol}	& \colhead{Value} & \colhead{Reference}}
    \startdata
    Total number of filaments for full-sky & $N_{\rm fil}$ & 180.5 million & Section~\ref{sec:MFs} \\
    Filament density & $n_{\rm fil}$ & 3898\,deg$^{-2}\times [I_{\rm dust}/({\rm MJy\ sr}^{-1})]$ & Section~\ref{sec:MFs} \\
    Size of the box & $S$ & 400\,pc & Section~\ref{sec:large-scale-magnetic-field} \\
    Large-scale magnetic field model & & \citet{2012ApJ...757...14J,2012ApJ...761L..11J} & Section~\ref{sec:large-scale-magnetic-field} \\
    RMS of isotropic random magnetic field & RMS($\bm{H}$) & 3\,$\mu$G & Section~\ref{sec:code-magnetic-field}\\
    Random isotropic magnetic field power law & $P(k)$ & $\propto k^{-4}$ & Section~\ref{sec:code-magnetic-field} \\
    Multipole limit for very long filaments & $\ell_{\rm limit}$ & 50 & Section~\ref{sec:skipping-large-filaments} \\ 
    Minimum length of filaments & $L_a^{\rm min}$ & 0.04\,pc & Sections~\ref{sec:properties-distribution-correlations}, \ref{sec:sizes-filament-La} \\
    Filament length, Pareto distribution & $p(L_a)$ & $\propto L_a^{-2.445}$ & Sections~\ref{sec:ratios-method}, \ref{sec:sizes-filament-La} \\
    Filament axis ratio & $\epsilon$ & $0.16 (L_a/L_a^{\rm min})^{+0.122}$ & Section~\ref{sec:ratios-method},  \\
    Filament misalignment angle dispersion & RMS($\theta_{LH}$) & 10\degree & Section~\ref{sec:orientation-filaments} \\
    Polarization fraction geometric dependence & $f_{\rm pol}$ & $\propto (L_a/L_a^{\rm min})^{-0.1}$ & Sections~\ref{sec:ratios-method}, \ref{sec:pol-fraction} \\
    Dispersion MBB SED & $\sigma_{\rho}$ & 0.15 & Section~\ref{sec:sed_method}, eq.~\ref{eq:perturbation}\\
    \enddata
\end{deluxetable*}

Most of the details of the geometric description of how we define filaments are in Appendix~\ref{sec:appendix-method}, while in the following subsections, we describe how our model is fine-tuned to match the Planck thermal dust spectra from \citet{planck_2018_xi}. A summary of the parameters used to define our model is given in Table~\ref{table:parameters}.

\begin{figure}
    \centering
    \includegraphics[width=1.0\columnwidth]{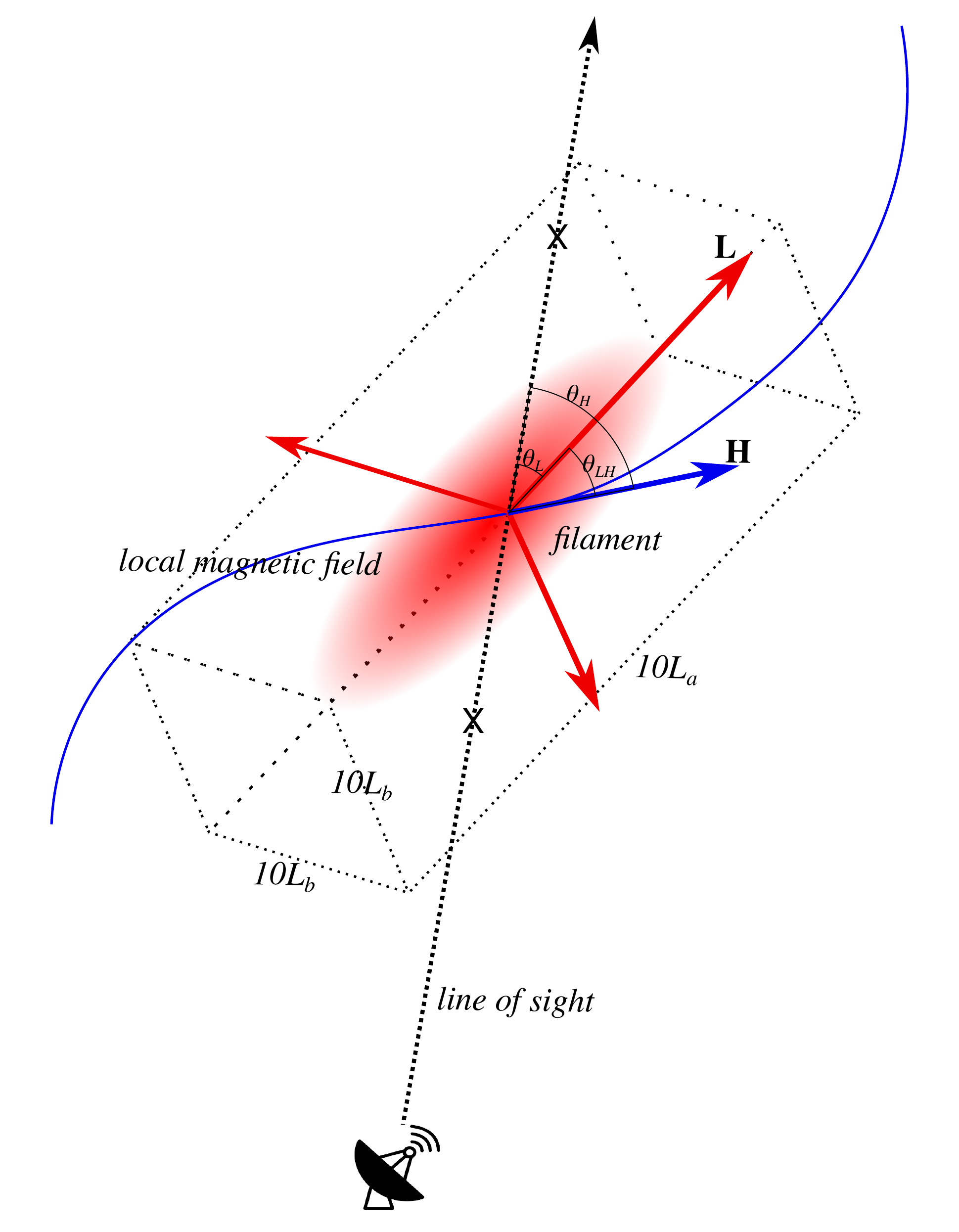}
    \caption{Orientation of a filament with respect to the magnetic field and the bounding box for our LOS integration. The filament long axis $\bm{L}$ is correlated but not perfectly aligned to the local magnetic field $\bm{H}$ as described in the main text. We integrate along the LOS between the two crosses, where the LOS intersects the filament's rectangular box.}
    \label{fig:sketch}
\end{figure}


\subsection{Filament properties, spatial distribution, and correlations} \label{sec:properties-distribution-correlations}


The small-scale power spectra derive from correlations of filaments with themselves, a one-filament contribution that corresponds to the one-halo term in the cosmological halo model \citep{1991ApJ...381..349S, 2000MNRAS.318..203S}.
As in \citet{filament_paper}, we model our filaments as prolate spheroids, each with a long semiaxis $L_{\rm a}$ and two short semiaxes $L_{\rm b}$. We model the density profiles as Gaussians.  The slopes of the power spectra are determined by the dependence on filament size of the halo abundance and properties.   The semi-major axis length of the filaments $L_a$ is drawn from a Pareto distribution $p(L_a) \propto L_a^{-\eta_L}$ and starting at a minimum length $L_a^{\rm min}$. We use $\epsilon = L_b/L_a$ as the axis ratio determining the shape of the prolate spheroid, which varies slightly with length.  The central densities of the filaments follow an empirical size relationship from the ISM, $n_0 \propto L_a^{-1.1}$ \citep{1981MNRAS.194..809L}. Since these relations are all power laws, the small-scale power spectra will also be power laws \citep{filament_paper}.

Our choice for how we place the filaments inside the cubic volume controls the correlations between filaments, providing a two-filament contribution that determines the large-scale power spectrum, unlike \citet{filament_paper}, who considered filaments on a single shell and only treated the one-filament term.
To reproduce the overall distribution of dust, we allocate our filaments across the sky according to a full-sky dust template $I_{\rm dust}$, in this case, the GNILC dust template from Planck \citep{2020A&A...641A...4P}, using a random Poisson distribution with an expected number of filaments per pixel $p$,
\begin{equation} \label{eq:lambda}
    \lambda(p) = \frac{N_{\rm fil} I_{\rm dust}(p)}{\sum_{p'} I_{\rm dust}(p') } \text{,}
\end{equation}
where $N_{\rm fil}$ is the total number of filaments. 
The filaments are given random radial positions but distributed so that the volume density of the filaments is constant along each LOS.

Filaments that subtend large angles can be generated by chance if they are very close or have a large intrinsic size (or may intersect the observer). We skip them when their angular size is larger than some limit multipole $\ell_{\rm limit}$.
We adopt a value of $\ell_{\rm limit} = 50$, which is equivalent to an angle $\sim 3.6$\degree, which proves to be a good threshold for leaving out the unrealistic very long filaments that would show up prominently in a $T$ map.

The magnetic field provides the two-filament correlations between the orientations of the filaments, as well as the orientation of the dust polarization.  We model the large-scale Galactic field, as well as a small-scale isotropic random field, where we generate a random correlated vector field inside the cube following a power spectrum. The details are in Appendix \ref{sec:magnetic-field}. We fix the RMS of the random isotropic component to be 3\,$\mu$G \citep{2008A&A...477..573S,2010MNRAS.401.1013J}, which is larger than the RMS of the large-scale model, which is $\sim 0.5$\,$\mu$G in the box we consider. Because of this, the large-scale Galactic magnetic field model choice has a very small effect. (We discuss possible modeling improvements in Section~\ref{sec:conclusions}.) We orient the filaments roughly following the direction of the local magnetic field, as in Fig.~\ref{fig:sketch}.  The filament long axis $\bm{L}$ is rotated by an angle $\theta_{LH}$ away from the local magnetic field $\bm{H}$. This angle is drawn from a Gaussian distribution with zero mean and standard deviation RMS($\theta_{LH}$). Then, to randomize the filament orientations, we rotate the $\bm{L}$ vector once more around the local magnetic field $\bm{H}$ by a random angle $\phi_{LH} \sim \unif(0,2\pi)$. Consistent with MHD simulations, we have used a $k^{-4}$ power spectrum for the generation of the random magnetic field. Large scales have much more power than small scales, and this results in large-scale coherent orientations of the magnetic field.  By sample variance, the orientation of these directions with respect to the galactic plane depends on the random seed for the field in our code.  The large-scale coherence affects the relative power of temperature and polarization fluctuations in the map and their cross-correlation.  Such an effect is absent for incoherent magnetic field directions (i.e. white power spectrum for the magnetic field).

\subsection{LOS integration}
For each filament, we integrate the LOSs that correspond to the individual pixels of a full-sky \textsc{healpix} map of a given resolution, $N_{\rm side} = 2048$ in our main case, projecting the image of the 3D filament onto the 2D surface of the celestial sphere (Appendix \ref{sec:los-integration}). Summing all the filaments in the population renders the full-sky image viewed by an observer located at the center of the cube.  As in Fig.~\ref{fig:sketch}, for integration, the profile is defined inside a rectangular box with a long side $10 L_a$ and the two short sides $10L_b$. 

Since most of the filaments will have a very small angular size, we would waste resources by sampling all filaments with the same resolution, where some of them would be sampled by several million pixels and others by a handful of pixels. To avoid this, we implement a mechanism to sample each filament with a variable resolution $N_{\rm side}^{\rm variable}$, determined by the filament size, which may be coarser or finer than our final resolution.
We then smooth or degrade to achieve the final resolution while avoiding pixel artifacts (Appendix~\ref{sec:variable-sampling}).

Finally, we extrapolate this map at specified frequencies, which can be done with a simple spectral energy distribution (SED), or an elaborate method to create some level of frequency decorrelation.

\subsection{Reproducing power spectrum ratios and slopes } \label{sec:ratios-method}


We aim to reproduce the power spectrum ratios of the EE, BB and TE spectra from \citet{planck_2018_xi}, as well as the slopes  $\alpha_{\rm XY}$.
To reproduce the $\mathcal{D}_{\ell}^{\rm BB}/\mathcal{D}_{\ell}^{\rm EE}$ ratio, we set the filament misalignment to the magnetic field RMS($\theta_{LH}$) and the axis ratio $\epsilon$.  \citet{filament_paper} noted that a tighter alignment between the magnetic field and filament axis leads to an excess of E power over B power (here Appendix~\ref{sec:single-filament-analysis} explores this idea in more detail). A smaller axis ratio $\epsilon$ (meaning thinner filaments) also increases the relative power of the E-modes over the B-modes.    \citet{filament_paper} incorrectly concluded that  there was a unique combination of misalignment and axis ratio that simultaneously fit the $E/B$ power ratio and the $r^{\rm TE}$ correlation when the polarization fraction was common for all filaments.  That conclusion was due to a now-fixed bug that underestimated both the misalignment effect on the power ratio and the overall level of $r^{\rm TE}$.  Here we find that the parameters RMS$(\theta_{LH})=10$\degree\ and $\epsilon = 0.16$ (at the minimum filament size) work well, but find these parameters are not unique; a thinner filament could work if less aligned.  We chose a particular combination because it works and is computationally convenient; extremely thin filaments are difficult to represent with a small number of pixels.  We also found that variety in the polarization fraction per filament is necessary for $r^{\rm TE}$, discussed below.



The TT slope is affected by the filament length distribution with the probability density function $p(L_a) \propto L_a^{-\eta_L}$. The $\eta_L$ index of the Pareto size distribution will shift all slopes at the same time, so we fix it to $\eta_L = 2.445$ which (in combination with Larson's law for the density distribution) will enable the TT spectrum slope to match the measurement of $\alpha_{\rm TT} \sim -2.6$. 
As  detailed in \cite{filament_paper}, to achieve different slopes for the different TEB spectra, we must put a filament length dependence on the axis ratio $\epsilon(L_a)$ and polarization fraction $f_{\rm pol}(L_a)$.  A positive slope on the model $\epsilon(L_a) \propto L_a^{\eta_\epsilon}$ means large scales will have less EE power over BB power compared to small scales, making EE have a shallower spectrum. Adopting $\eta_\epsilon = 0.122$ creates a difference of $\alpha_{\rm BB} - \alpha_{\rm EE} \sim -0.11$ between the two slopes. Fig.~\ref{fig:ratios_planck_binning} compares data and model EE and BB ratios.

The difference between the temperature and polarization slopes is fixed by the length dependence of the polarization fraction with the model $f_{\rm pol} \propto L_a^{-\eta_{f_{\rm pol}}}$. We fix it to $\eta_{f_{\rm pol}}=0.1$, which shifts the slopes of the TE, EE and BB spectra to approximately their measured values. All of these values were found by running the semianalytic filament code from \cite{filament_paper} until we converge on satisfactory results. The significant digits on the slopes $\eta_\epsilon$, $\eta_{f_{\rm pol}}$ and $\eta_L$ are related to the sensitivity of the semianalytic code. For example, a $\pm0.01$ change in slope $\eta_\epsilon$ changes the EE and BB slopes by $\sim 0.03$. We aimed at matching the Planck-measured slopes within $\sim 0.01$ of their best-fit values, comparable to or smaller than the errors.


\subsection{Polarization fraction distribution} \label{sec:pol-fraction} 
\begin{figure*}
    \centering
    \includegraphics[width=1.0\textwidth]{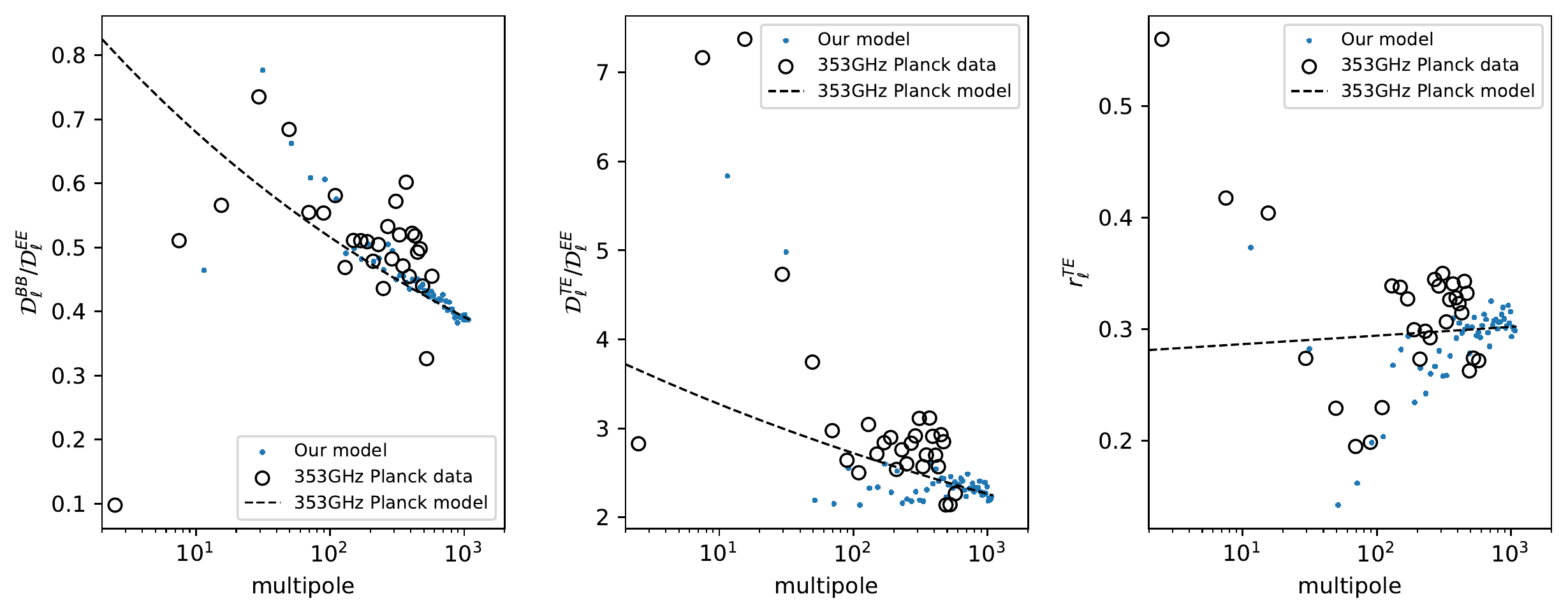}
    \caption{Moderate-resolution BB/EE, TE/EE and $r_{\ell}^{\rm TE}$ power spectrum ratios comparison between our filament model and the Planck 353\,GHz dust spectra from \citet{planck_2018_xi}. These are used to calibrate the $f_{\rm pol,0}$ distribution to match the Planck ratios, as explained in the text. The circles are the Planck data points, using their binning scheme $2\leq\ell<600$ and the LR71 mask ($N_{\rm side} = 512$). The blue lines are the ratios calculated from the power law models for each spectrum. The TT spectrum power-law fit ($\alpha_{\rm TT}=-2.60\pm0.03$) is calculated by this work, and the other three are fitted by \citet{planck_2018_xi}. The blue points are the ratios calculated from our filament model in the LR71 mask but with a binning scheme $2\leq\ell<1100$. Notice how our model matches the extrapolated ratios beyond $\ell \geq 600$. 
    }
    \label{fig:ratios_planck_binning}
\end{figure*}

The polarization fraction $f_{\rm pol}$ dictates the relative strength of polarization with respect to intensity. It has a geometric dependence $\propto \sin^2\theta_H$ \citep[][where $\theta_H$ is the angle between the LOS and the local magnetic field]{2000ApJ...544..830F} and some normalization constant $f_{\rm pol,0}$. As mentioned above, we include a slight power law dependence on the filament length, $f_{\rm pol} \propto L_a^{-\eta_{f_{\rm pol}}}$. Then, the polarization fraction that multiplies $Q,U$ is $f_{\rm pol} \propto f_{\rm pol,0} L_a^{-\eta_{f_{\rm pol}}} \sin^2(\theta_H)$. As noted in \cite{filament_paper}, $r_{\ell}^{\rm TE}$ and $\mathcal{D}_{\ell}^{\rm TE}/\mathcal{D}_{\ell}^{\rm EE}$ depend on the $f_{\rm pol,0}$ distribution as follows:
\begin{align}
    r_{\ell}^{\rm TE} &\propto \langle f_{\rm pol,0} \rangle / \langle f_{\rm pol,0}^2 \rangle \\
    \mathcal{D}_{\ell}^{\rm TE}/\mathcal{D}_{\ell}^{\rm EE} &\propto \langle f_{\rm pol,0} \rangle / \langle f_{\rm pol,0}^2 \rangle ^{1/2} \text{.}
\end{align}
This means that we need an $f_{\rm pol,0}$ distribution with a domain limited to $[0,1]$ and convenient control over the mean and variance. The beta distribution fulfills these requirements.  It depends on two parameters, $\alpha$ and $\beta$, which together determine the mean and variance of the distribution.  We sample with
\begin{equation} p(f_{\rm pol,0}) \propto {\rm PDF_{Beta}}(\alpha,\beta)\end{equation}
and we can calibrate $\alpha$ and $\beta$ to achieve a specific relation between the mean and the variance and increase or decrease $r_{\ell}^{\rm TE}$ and $\mathcal{D}_{\ell}^{\rm TE}/\mathcal{D}_{\ell}^{\rm EE}$ as needed to fit the Planck-modeled ratios. 

As mentioned in Section~\ref{sec:properties-distribution-correlations}, the coherent orientation of the random magnetic field will change slightly with the seed used to generate it. The polarization fraction calibrates the ratio between temperature and polarization, so in order to match the Planck observations, a new polarization fraction calibration is needed when changing the magnetic field seed. In practice, this means that to match the Planck spectra, the $\alpha$ and $\beta$ parameters of the beta distribution will be different for each magnetic field seed.  Physically, this also means that the temperature-to-polarization relationships seen in foregrounds are likely not universal but rather depend on the local magnetic field structure. From a different location in the Milky Way or from inside an analogous galaxy, an observer would see a different realized magnetic field, altering the ratio between the temperature and polarization of dust.


These polarization fraction-dependent quantities are also illustrated in Fig.~\ref{fig:ratios_planck_binning}, where we plot the Planck-measured ratios at 353\,GHz from \citet{planck_2018_xi} (black circles) and the power law models (black dashed lines; the TT spectrum is fitted by this work, and the other three are fitted by the Planck team). The ratios from our filament model are calculated up to $\ell < 1100$, shown as blue points.  We tune the beta distribution parameters such that the ratios from our filament model fit the Planck-modeled power law ratios.

\subsection{Normalization} \label{sec:normalization}

\begin{figure}
    \centering
    \includegraphics[width=1\columnwidth]{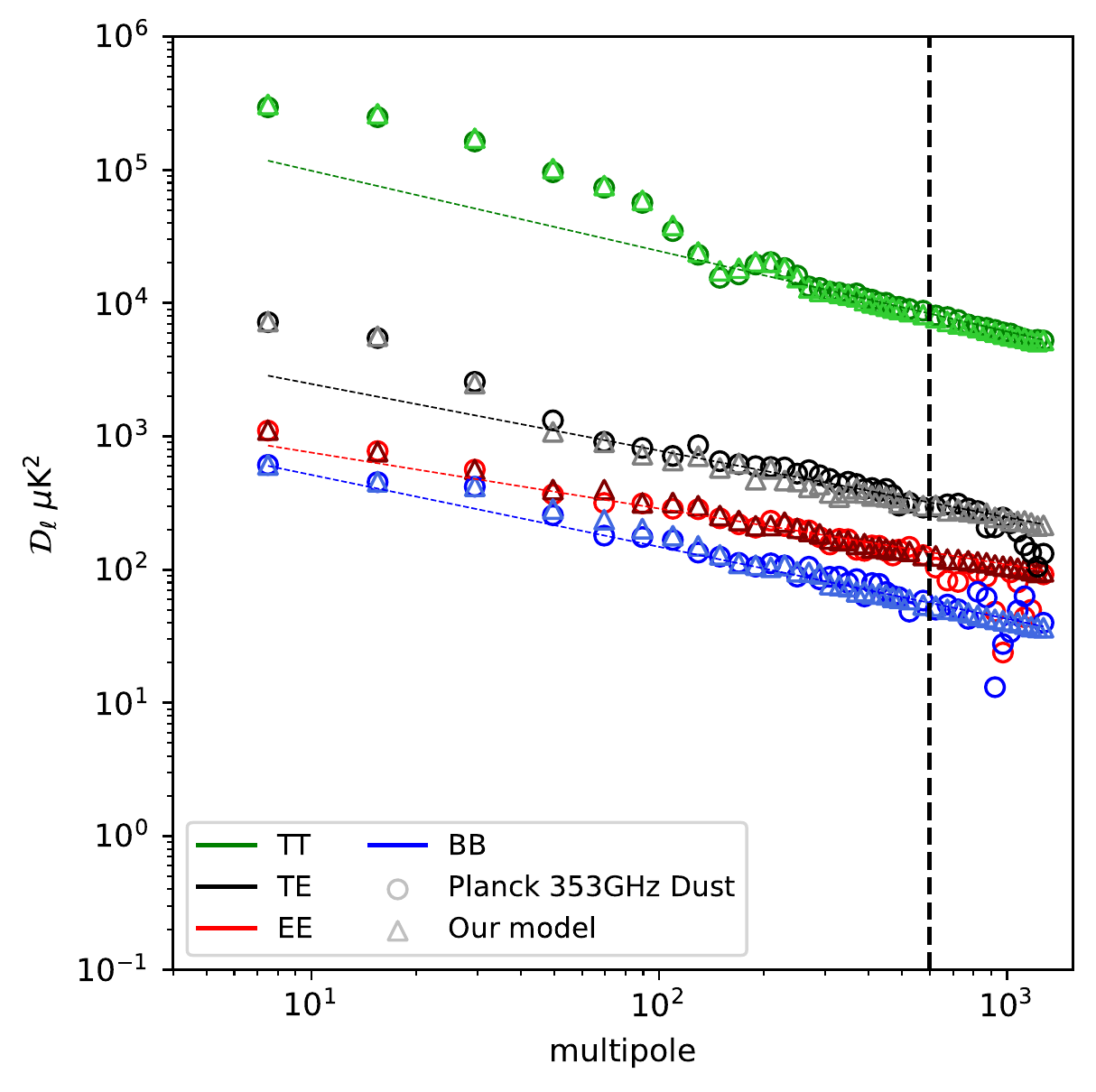}
    \caption{Moderate-resolution power spectrum comparison between our filament model (triangles) and the Planck 353\,GHz dust spectra from \citet[][circles]{planck_2018_xi}, using their binning scheme $2\leq\ell<600$ and the LR71 mask ($N_{\rm side} = 512$). Here TT is green, TE is black, EE is red, and BB is blue. The dashed vertical line is the limit $\ell = 600$. In the range $600 \leq \ell < 1300 $, we use bins with $\Delta \ell = 50$. The $Q$ and $U$ maps in our model, and therefore the polarization spectra, have their large-scale emission filled using the Planck frequency maps, as explained in the text. The dashed lines are the power law model for each spectrum. The TT is fitted by this work ($\alpha_{\rm TT}=-2.60\pm0.03$), and the other three are fitted by \citet{planck_2018_xi}.}
    \label{fig:example-filling-large-scale}
\end{figure}

The different parts of the model scale differently with the mean number density of the filaments.  Before normalization, the large-scale, two-filament contribution to the power spectrum scales like $n_{\rm fil}^2$, while the small-scale, one-filament contribution scales like $n_{\rm fil}$ \citep{1991ApJ...381..349S}.  Both the one- and two-filament terms scale as the square of the normalization for the filament mass density profile.  We have to adjust both of these parameters to match the Planck power spectra on large and small scales.
In practice, we choose our overall temperature-unit normalization to fix the small-scale polarization power spectra on small scales in all cases, since that is the quantity of most interest.  The resulting power spectrum normalization scales as $1/n_{\rm fil}$ to counteract the dependence of the smale-scale, one-filament term.  Then we can examine the large scales to deduce the $n_{\rm fil}$ that does not under- or overproduce the large-scale power in the two-filament term.

Because of the tuning of the polarization fraction, the Stokes parameter maps are calibrated among each other.
We use the following procedure to calibrate to $\mu{\rm K}$, based on the 353 GHz dust \citet{planck_2018_xi} EE and BB polarization spectra calculated 
for the LR71 mask, listed in their Table C.1 with error bars from simulations,
in bins over the range $2 \leq \ell < 600$.  We need to restrict this multipole range to examine only the one-filament contribution and fit with a standard $\chi^2$ estimator for the spectrum's amplitude.  The best reduced $\chi^2$ is achieved for $280 < \ell < 600$.  We compute the EE and BB spectra of our filament model in the same LR71 mask.  After this procedure, our filament model will have Stokes parameter maps in physical units.
We then try several values for the filament density $n_{\rm fil}$, which, in practice, is implemented by setting a total number of filaments $N_{\rm fil}$\footnote{The $N_{\rm fil}$ is dependent on the sky fraction or mask considered, while the filament density $n_{\rm fil}$ in units of deg$^{-2}\times [I_{\rm dust}/({\rm MJy\ sr}^{-1})]$ is independent of this. The $N_{\rm fil}$ will be different for a full-sky versus partial-sky simulation, but $n_{\rm fil}$ will be the same.}. Using $N_{\rm fil} = 180.5 \times 10^6$ filaments for the full-sky (which is a filament density $n_{\rm fil}=3898$\,deg$^{-2}\times [I_{\rm dust}/({\rm MJy\ sr}^{-1})]$, where $I^{\rm dust}$ is the dust intensity at any given pixel at 353\,GHz), we can produce the large-scale power coming from the two-filament term that matches the Planck 353\,GHz $\mathcal{D}_{\ell}^{\rm TT}$ spectrum. Using the Knox formula to estimate the error bars, we find $\chi^2=269$ (for 29 multipole bins) for the $\mathcal{D}_{\ell}^{\rm TT}$ spectrum fit using the \citet{planck_2018_xi} binning scheme in the range $2\leq\ell<600$. The TT spectrum comparison between the Planck dust 353\,GHz emission and our filament model is shown in Fig.~\ref{fig:example-filling-large-scale} (green circles and triangles). 

As mentioned in Section~\ref{sec:pol-fraction}, changing the seed of the random magnetic field alters the temperature-to-polarization relationships.  Since we calibrate the $\mu$K units with respect to the EE and BB spectra, the TT and TE spectra will change with respect to the Planck-measured spectra. The standard deviation of the polarization-to-TT calibration is $\sim 10$\% when changing the magnetic field seed but keeping the same (untuned) polarization fraction distribution.

\subsection{Large-scale polarization template} \label{sec:fill-large-scale}

\begin{figure}
    \centering
    \includegraphics[width=1\columnwidth]{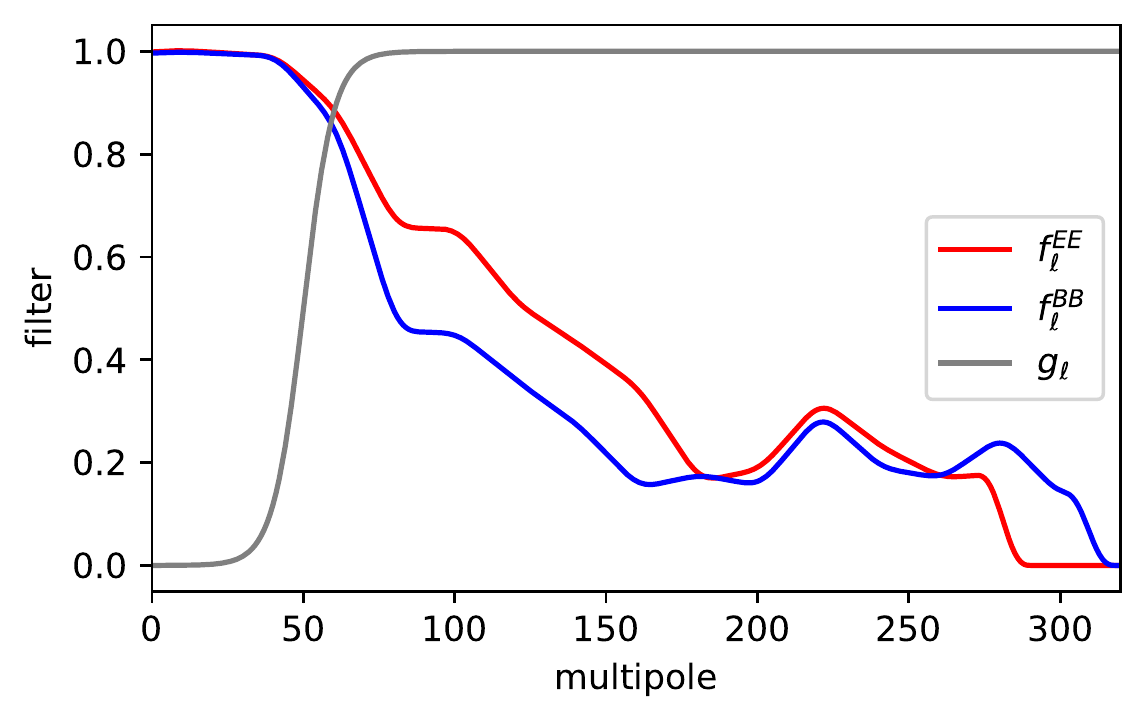}
    \caption{Filters for blending the filament model E and B fields with a Planck template of the real sky, as defined in eq.~\ref{eq:template-filter}. Note that at scales $\ell < 50$, all emission is contributed by the Planck template, while at higher multipoles, the filament model comes to dominate the mixture.}
    \label{fig:ad hoc-filter}
\end{figure}

Although we can reasonably approximate the large-scale temperature map by having the filament density trace the dust in the Milky Way, we do not reproduce the large-scale polarization. Our 3D model of the dust distribution and Galactic magnetic field is insufficient to do so, as it does not include the specific features that are crucial to reproducing the large-scale $Q$ and $U$ maps. 

To address this, we can replace the large scales of the $Q$ and $U$ maps with a polarized dust template of the real sky to create a hybrid model. 
First, we suppress the large scales $\ell < 50$ produced in our filament model by using a logistic function as a spherical harmonic high-pass filter $g_{\ell}$. We want to match the target thermal dust spectra calculated by \citet{planck_2018_xi}, by filling the difference at large scales with the power spectra of a template map. This map, which in our case is the 353\,GHz full mission map from Planck DR3 with the SMICA CMB map subtracted \citep{2020A&A...641A...4P}, must be filtered by the ad hoc spherical harmonic filter $f_{\ell}^{XX}$ such that
\begin{equation} \label{eq:template-filter}
    \mathcal{D}_{\ell}^{XX, \rm target} = (f_{\ell}^{XX})^2 \mathcal{D}_{\ell}^{XX, \rm template} + g_{\ell}^2 \mathcal{D}_{\ell}^{XX, \rm filaments} \text{,}
\end{equation}
where $X \in E,B$, $\mathcal{D}_{\ell}^{XX, \rm filaments}$ are the spectra of our filament model, and $\mathcal{D}_{\ell}^{XX, \rm target}$ is the dust spectrum we want to match, as calculated by \citet{planck_2018_xi}, crossing the two half-missions and subtracting the best-fit CMB, as described in Section~\ref{sec:data}. This filter, $f_{\ell}^{XX}$, is calculated with $\Delta \ell = 20$ bins, and it is smoothed with a Hamming window to avoid sharp edges and ringing effects. Also, we force $f_{\ell}^{XX} = 0$ when $\ell \geq 300$ (since at these scales, we want the whole emission to come completely from our model) or is undefined, i.e. $ \mathcal{D}_{\ell}^{XX, \rm target} < \mathcal{D}_{\ell}^{XX, \rm filaments} $. We show the ad hoc filter $f_{\ell}^{XX}$ and high-pass filter $g_{\ell}$ in Fig.~\ref{fig:ad hoc-filter}.

Then, the $Q$ and $U$ \emph{all-scale} hybrid map in our filament model is the sum of the filtered \emph{template} map plus the high-pass filtered \emph{small-scale} map (our filament model), following eq.~\ref{eq:template-filter}. The resulting all-scale power spectra from our filament model are shown in Fig.~\ref{fig:example-filling-large-scale}, compared to the target dust spectra we want to match. All spectra use the LR71 mask. We also include the TT and TE spectra calculated with our filament model $T$ map, which has no large-scale filling, but still it is able to reproduce the Planck-measured spectra fairly accurately, as will be described in Section~\ref{sec:MFs}.

In the figure, we show the \citet{planck_2018_xi} spectra compared to our filament model spectra in the same binning scheme, $2\leq\ell<600$. We also extend the bins $600 \leq \ell < 1300$ with size $\Delta \ell = 50$, showing the consistent power law emission of our model, while the Planck polarization emission has a very low signal-to-noise ratio and the bandpowers are noisy. 

\begin{figure*}
    \centering
    \includegraphics[width=1.0\textwidth]{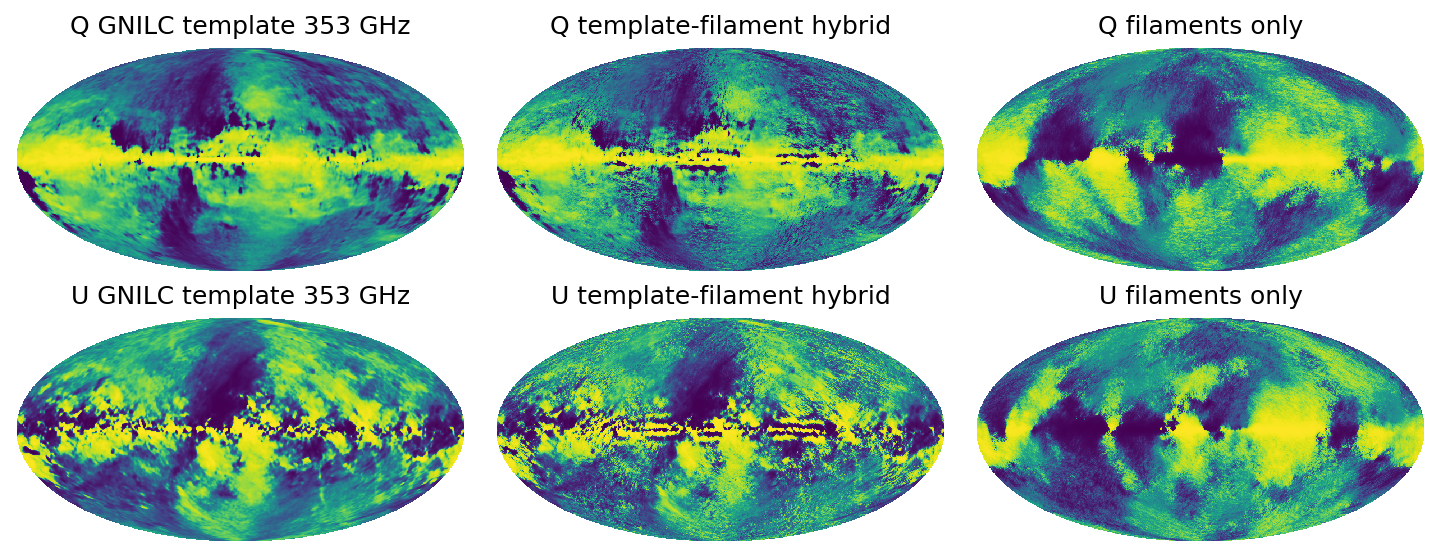}
    \caption{full-sky $Q$ and $U$ emission from Planck's GNILC dust template at 353\,GHz (left column), a hybrid filament model (middle column) that replaces the large scales with Planck's template, and our raw filament-only model (right column). The hybrid model (middle column) shows stripe artifacts produced by the high-pass harmonic filtering along the Galactic plane region and should not be used there.
    }
    \label{fig:qu_fullsky}
\end{figure*}

Because the Galactic plane emission is very bright in $Q$ and $U$, filtering our filament model with the high-pass harmonic filter $g_{\ell}$ produces very prominent stripe artifacts near the galactic plane. These stripes are visible if we view the full-sky, unmasked map of our filament model (see Fig.~\ref{fig:qu_fullsky}). Therefore, in the hybrid model, we exclude the filament model along the plane, keeping the sky in the Planck DR2 Galactic mask with $f_{\rm sky}=0.9$. We note that the Galactic plane $Q$ and $U$ emission from our final model will not contain small-scale emission, and we warn the reader to not use the polarization from our model inside the Galactic plane.

\subsection{Spectral Energy Distribution} \label{sec:sed_method} 

With our model, we aim to produce multifrequency simulations of the thermal dust emission at any arbitrary frequency channel. We can start with the usual MBB SED used to model the thermal dust emission, given by 
\begin{equation} \label{eq:MBB}
    S_{\rm dust}(\nu,\beta_{\rm dust},T_{\rm dust}) \propto \nu^{\beta_{\rm dust}+3} / [ \exp(h\nu/kT_{\rm dust}) - 1 ]
\end{equation}
in surface brightness units, where $\nu$ is the frequency; $h$ and $k$ are the Planck and Boltzmann constants, respectively; and $\beta_{\rm dust}$ and $T_{\rm dust}$ are the free spectral parameters of emissitivity index and dust temperature, respectively. When a full-sky map is generated at some frequency (e.g. 353\,GHz), it is straightforward to multiply this map with an MBB spectral law at chosen frequencies with either spatially constant or variable dust spectral parameters. In this case, we only need to generate a single template of thermal dust Stokes parameters at an anchor frequency, and the extrapolation to other frequencies can be done separately.

Another option is to introduce frequency decorrelation, where the flux measured between two or more frequencies is not a constant factor. \citet{2017A&A...599A..51P,planck_2018_xi} measured this on different Galactic masks with different sky fractions. Recently, \citet{2021A&A...647A..16P} tried to measure the frequency decorrelation on individual LOSs within the Galactic poles areas using the Planck maps, as well as the 3D information along the LOS provided by H I spectral observations. They modeled the ratio between the 217 and 353\,GHz thermal dust flux as some constant $\delta$ that is perturbed by a small Gaussian random variable $\rho$ with zero mean and standard deviation $\sigma_{\rho}$, given by 
\begin{equation} \label{eq:perturbation}
    \frac{S_{\rm dust}(217,\beta_{\rm dust},T_{\rm dust})}{S_{\rm dust}(353,\beta_{\rm dust},T_{\rm dust})} = \delta (1 + \rho) \text{.}
\end{equation}
The $\delta$ flux ratio represents the mean ratio along each LOS, which can be calculated with the best-fit $\beta_{\rm dust}$ and $T_{\rm dust}$ parameters from the Planck GNILC estimation, as described in Section~\ref{sec:data}. The $\sigma_{\rho}$ standard deviation represents the degree of random variability in the dust SED along the LOS, which can be accomplished by randomizing $\beta_{\rm dust}$ and $T_{\rm dust}$, adding new parameters, or even replacing the spectral model completely. 

In our case, we model the frequency decorrelation by generating a random dust MBB SED for each individual filament. Then, since our model is the addition of millions of maps of individual filaments, we naturally create a way to decorrelate different frequencies. We generate the random $\beta_{\rm dust}$ index, and we fix the $T_{\rm dust}$ parameter to the best-fit value found by Planck on each LOS. Since the 217 and 353\,GHz frequencies are within the Rayleigh-Jeans area of the MBB, the impact of varying $T_{\rm dust}$ is limited. We choose to put all the dust SED variability in the $\beta_{\rm dust}$ index. Using eq.~\ref{eq:MBB} and inverting eq.~\ref{eq:perturbation} for $\beta_{\rm dust}$, we find
\begin{equation} \label{eq:beta-dust}
    \beta_{\rm dust} = \log \left[ \delta (1+\rho) \frac{e^{217{\rm GHz}h/kT_{\rm dust}}-1}{e^{353{\rm GHz}h/kT_{\rm dust}}-1} \right] /\log(217/353) - 3 \text{,}
\end{equation}
where the $T_{\rm dust}$ temperature and the ratio $\delta$ are coordinate-dependant, and consequently, the random $\beta_{\rm dust}$ will also be. 

Also, as we explained in Section~\ref{sec:fill-large-scale}, we fill the large-scale emission of the $Q$ and $U$ maps with the Planck 353\,GHz frequency map, which means that we need some recipe to fill those scales at any arbitrary frequency. The procedure to do this is the following. We take the large-scale fill-in map at 353\,GHz (which has been filtered in harmonic space by the ad hoc filter in eq.~\ref{eq:template-filter}), and we extrapolate it to the desired frequency using a regular MBB from eq.~\ref{eq:MBB}. We use the best-fit maps of $\beta_{\rm dust}$ and $T_{\rm dust}$ parameters from the GNILC estimation. 
Our model does not include polarization frequency decorrelation on scales where the large-scale fill-in contributes most of the emission, i.e. $\ell \lesssim 50$. In the transition scales, $\ell = 50-300$, there will be some level of frequency decorrelation, since a fraction of the emission is contributed by our filament model.
We have checked that this works well at 217\,GHz and frequencies relevant for the dust emission by comparing it with the Planck frequency map, finding a smooth transition between the large- and small-scale angular power spectra despite being derived completely separately. At lower frequencies, $\lesssim 50$\,GHz, the mismatch between the perfect MBB law at large scales and the decorrelated MBB law at small scales is enough to create a break in the power spectrum, although in this frequency range, dust is a minor foreground. 

Our filament model will contain a polarization frequency decorrelation at $\ell \sim 80$, where the recombination bump in the CMB primordial BB spectrum is located. It has less decorrelation than at $\ell \geq 300$, because at $\ell \sim 80$, the model is a mixture of the fill-in template and our filament model.

\section{Results} \label{sec:results}

In the following subsections, we detail the results of our filament model and how they match the Planck results. We summarize the model parameters we use in Table~\ref{table:parameters}. (We do not list the parameters for the polarization fraction beta distribution in the table because, as explained in Section~\ref{sec:pol-fraction}, to match Planck, they vary with the particular seed we used to generate the random magnetic field.  For our particular case, the values were $\alpha=0.07734$ and $\beta=0.37448$.)

\subsection{Maps}  \label{sec:maps}

\begin{figure*}
    \centering
    \includegraphics[width=1.0\textwidth]{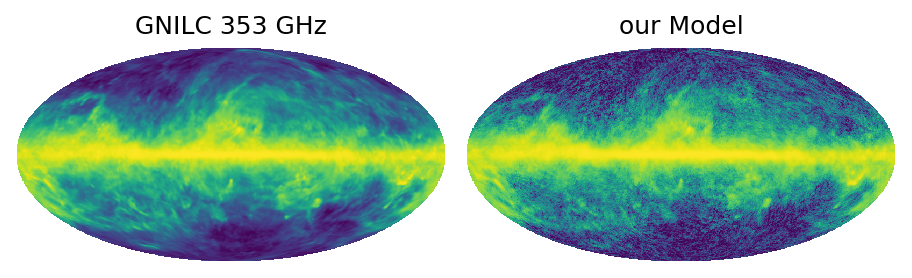}
    \includegraphics[width=.85\textwidth]{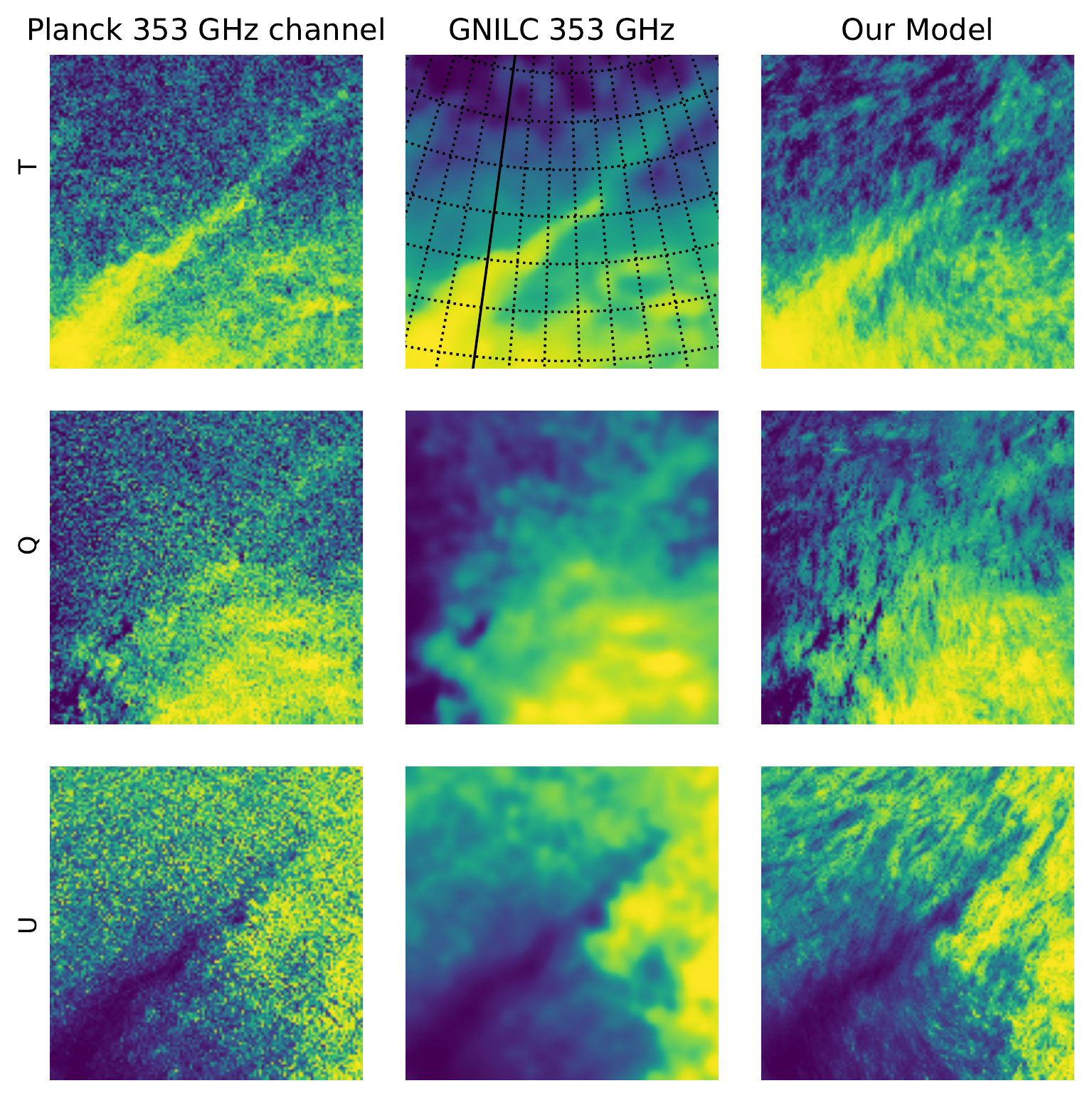}
    \caption{Top: temperature full-sky comparison between the GNILC dust template at 353 GHz from \citet{2020A&A...641A...4P} (left) and our simulated map (right). Bottom: close-up comparison in a $30\times30$\,deg patch centered at $l=50$\degree, $b=-10$\degree. We compare the Planck 353\,GHz channel (left column, with a resolution of $\sim 5\arcmin$), the GNILC dust template (middle column, with a resolution of $80\arcmin$), and our filament model (right column). The three rows are TQU. The Planck 353\,GHz Q and U maps are downgraded to $N_{\rm side}=256$ to average out the noise. 
    }
    \label{fig:tqu-maps}
\end{figure*}


In Fig.~\ref{fig:qu_fullsky} we show the full-sky polarization maps.  
In Fig.~\ref{fig:tqu-maps} (top), we show the full-sky comparison in temperature between our filament model and the GNILC dust template. Our temperature map comes only from the combined emission of millions of filaments stacked together. There are no data in it other than the GNILC template that modulates the probability to place the random filaments. In Fig.~\ref{fig:tqu-maps} (bottom), we show the zoomed $30\times30$\degree patch centered in the prominent superfilament north of the Ophiuchus region.
We compare the Planck 353\,GHz frequency channel (left column), which has a resolution of $\sim 5\arcmin$, the GNILC dust template (middle column); and our filament model (right column). The top, middle, and bottom rows are $T$, $Q$ and $U$. Our model is limited by the fact that it is composed of many small filaments that are oriented randomly with respect to their local magnetic field. Their centers might be in the correct place, but their orientations will not be correlated along tens of degrees into the shape of such a super dust filament. To achieve this, we would need a model of the Galactic magnetic field and dust distribution including all of the particular structures. In $Q$ and $U$, we fill the large-scale structure as described in Section~\ref{sec:fill-large-scale}. However, the extra small-scale detail that our filament model produces is clear in this comparison.

\begin{figure*}
    \centering
    \includegraphics[width=1.0\textwidth]{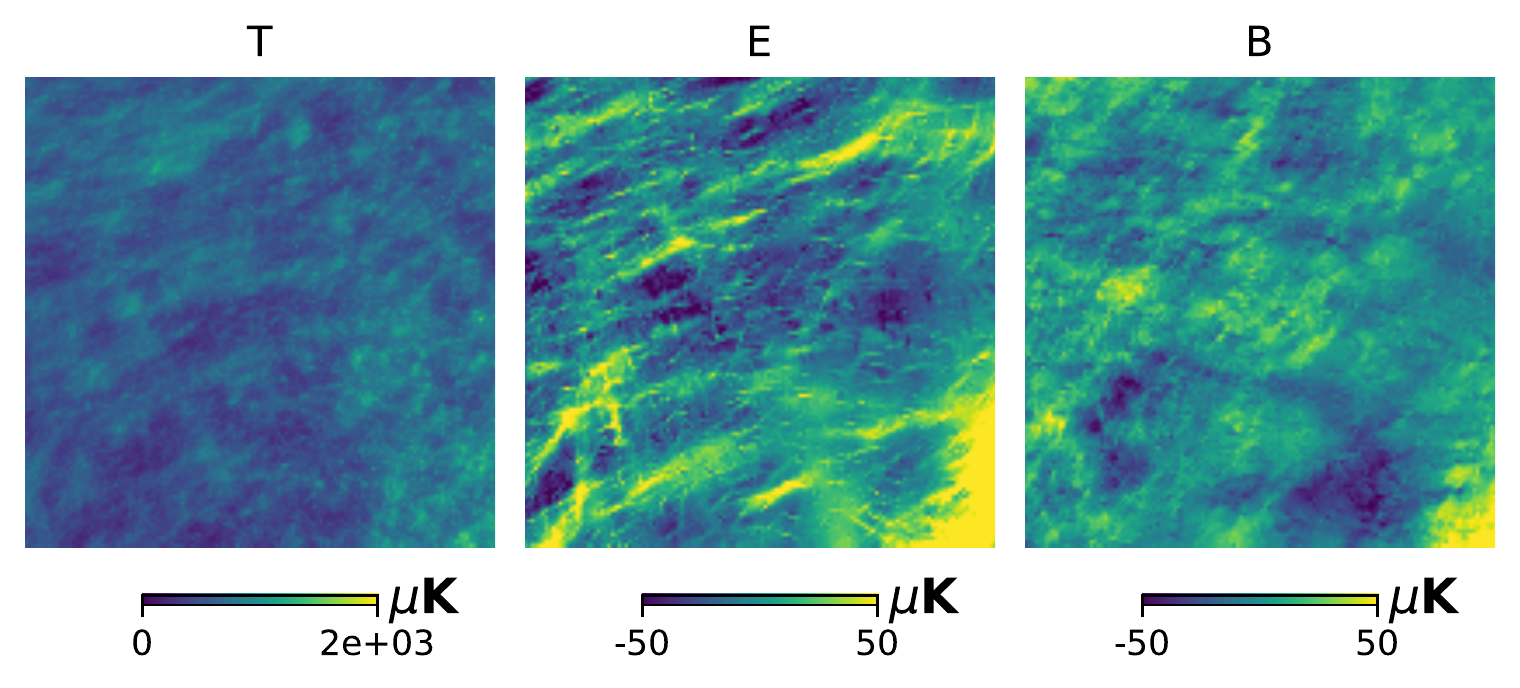}
    \caption{Close-up on a $10\times10$\degree patch centered at the north Galactic pole of our filament model. We show the TEB maps. The mean of each patch is subtracted from the E and B maps. 
    }
    \label{fig:TEB_northpole}
\end{figure*}

Fig.~\ref{fig:TEB_northpole} shows the zoomed-in patch centered in the north Galactic pole with a size of $10\times10$\degree for our filament model. We show the $T$ field, along with E and B. The groups of tiny filaments clumped along the magnetic field lines are clearly visible. We can see the E-mode domination over the B-mode; the E field runs positive along each filament axis, while the B field is a much weaker quadrupole pattern.  This is expected, given our strong alignment of the filaments (compare to Fig.~2 of \citet{filament_paper}).

\subsection{Power spectra and ratios} \label{sec:results-spectra}

\begin{figure}
    \centering
    \includegraphics[width=1.0\columnwidth]{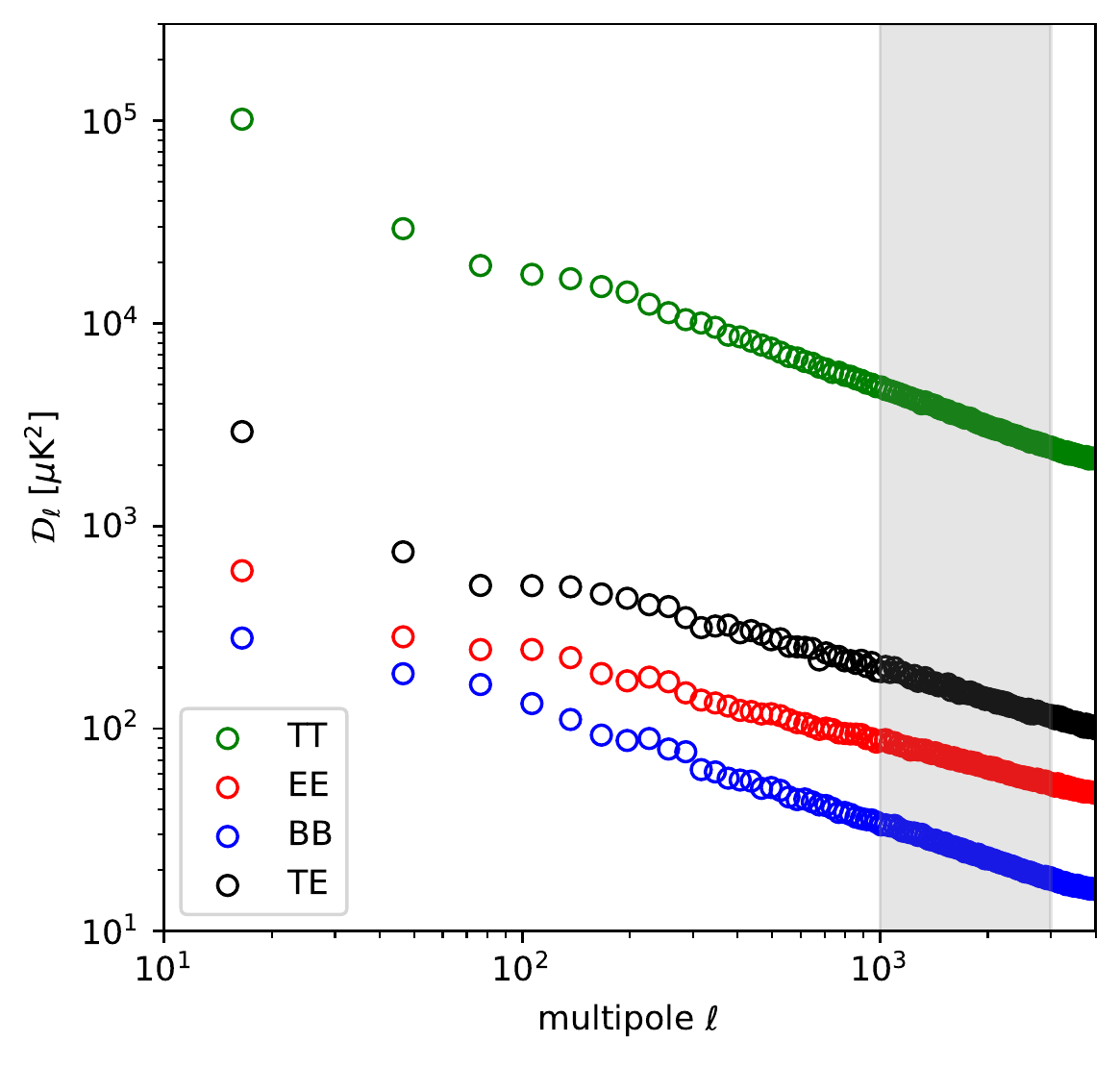}
    \caption{Higher-resolution power spectra from our filament model at 353\,GHz, calculated with the Planck DR2 Galactic mask with $f_{\rm sky}=0.7$ ($N_{\rm side}=2048$), apodized with a $2^{\circ}$ Gaussian kernel. We fit the slopes in the $1000 \leq \ell \leq 3000$ range (gray area). Note that our model has dust emission even at very small scales, $\ell \sim 4000$. 
    }
    \label{fig:Dell_mask2048}
\end{figure}

We present the power spectra of our filament model at 353\,GHz in Fig.~\ref{fig:Dell_mask2048}. These are calculated with the Planck DR2 Galactic mask ($f_{\rm sky}=0.7$), which is produced at $N_{\rm side}=2048$ natively. We calculate the power spectra up to $\ell_{\rm max}=6000$ with bins $\Delta \ell = 30$. The TB and EB spectra are consistent with zero. We emphasize the ability of our model to produce a consistent signal in the form of a power law down to very small scales.

We fit a power law to our filament model power spectra in the multipole range $1000 \leq \ell \leq 3000$ (gray area in the figure). We use the Knox formula to account only for sample variance in the error bars. As explained in Section~\ref{sec:ratios-method}, our filament model can produce different tunable different slopes for the different spectra. Our fit finds $\alpha_{\rm TT} = -2.633 \pm 0.003$, $\alpha_{\rm EE} = -2.459 \pm 0.004$, $\alpha_{\rm BB} = -2.590 \pm 0.003$, and $\alpha_{\rm TE} = -2.511 \pm 0.007$. We do not extend our fit to multipoles $\ell \gtrsim 2 N_{\rm side}$, since the very small filaments at these scales start approaching the point where they look like point sources.

We also fit the slopes at large/medium scales to compare directly with the \citet{planck_2018_xi} power law fit results. We show the spectra of our filament model in the Planck binning scheme ($2 \leq \ell < 600$) with the LR71 mask in Fig.~\ref{fig:example-filling-large-scale}. We fit the polarization spectra in the multipole range $40 \leq \ell < 600$ and using the same binning scheme as \citet{planck_2018_xi}. Our fit finds $\alpha_{\rm EE} = -2.50 \pm 0.02$, $\alpha_{\rm BB} = -2.65 \pm 0.02$, and $\alpha_{\rm TE} = -2.48 \pm 0.02$. 

We notice that these slopes are slightly different from the predictions (Section~\ref{sec:ratios-method}) made with the \citet{filament_paper} semianalytical code.  We attribute this difference to the fact that the semianalytical code assumes an isotropic distribution of the filaments and the magnetic field, while in our filament model, neither of these are true; the filaments are not isotropic, since they are arranged by the Galactic template, and the magnetic field does not have a white spectrum. Departing from these idealized conditions, we notice a slight steepening of the polarization spectra. Also, the fact that the error bars of the Planck fit contain both sample variance, and instrumental noise, while our filament model only includes sample variance should be taken into account. The slopes get closer to the target (Planck) values 
in the higher multipoles, $1000 \leq \ell \leq 3000$, since at these scales, the model is closer to the idealized, one-filament-dominated conditions than the semianalytical code models.

We also fit the TT spectrum for our filament model in the range $300 \leq \ell < 600$, finding $\alpha_{\rm TT}=-2.62 \pm 0.03$, which agrees with the value measured in the Planck map by ourselves. All of the fitted parameters for our filament model spectra are listed in Table~\ref{table:fitted_quantities}.

\begin{figure}
    \centering
    \includegraphics[width=1.0\columnwidth]{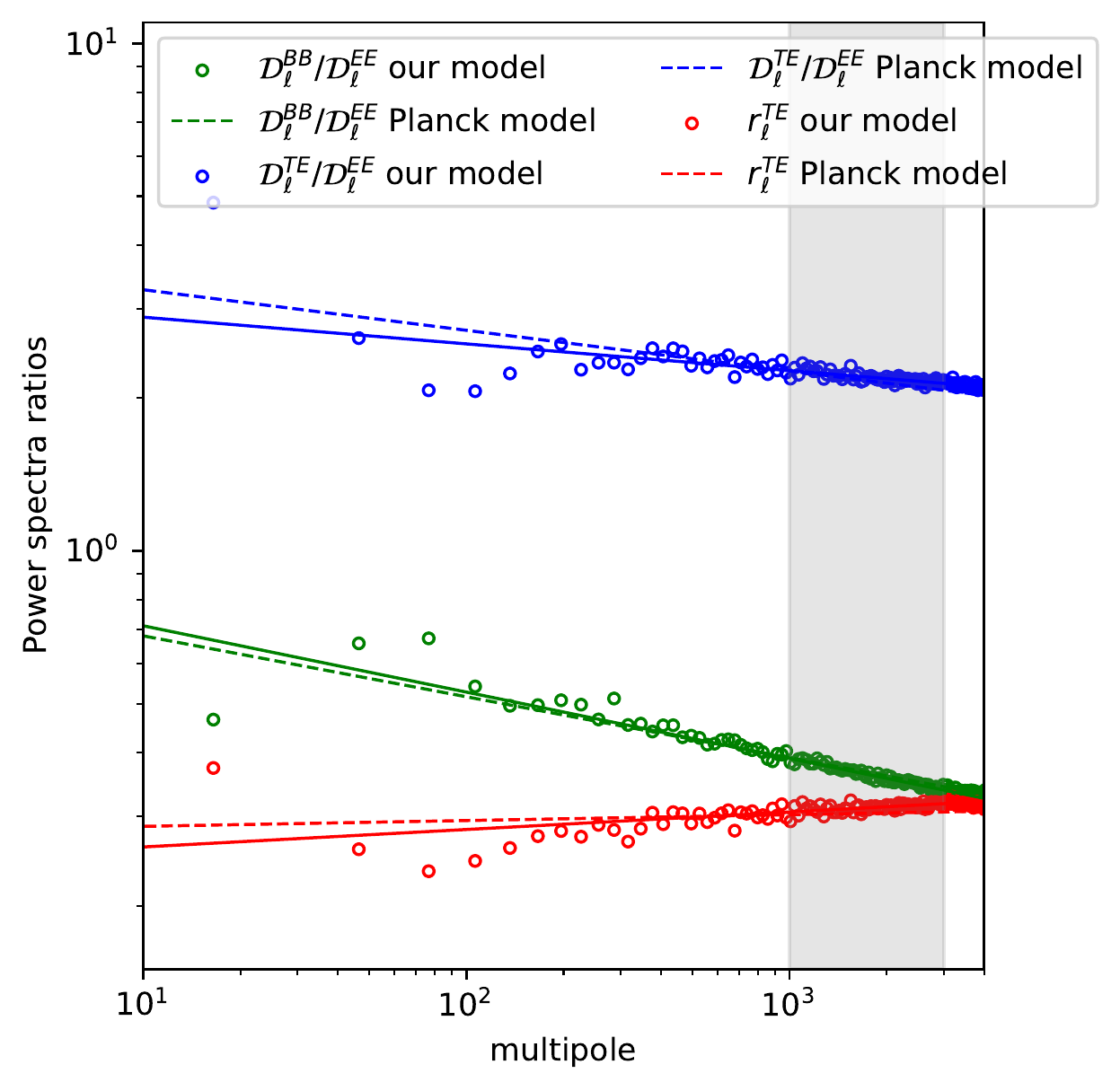}
    \caption{Higher-resolution power spectrum ratios from our filament model at 353\,GHz, calculated with the DR2 Galactic mask with $f_{\rm sky}=0.7$ ($N_{\rm side}=2048$), apodized with a $2^{\circ}$ Gaussian kernel. The circles are the ratios from our model, and the dashed lines are the targeted ratios modeled by the power law fit to each spectrum (extrapolated to very small scale) done in \citet{planck_2018_xi}, except for the TT spectrum, which is fitted by this work. The solid lines are the power laws fitted directly to the realization of our model and listed in Table~\ref{table:fitted_quantities}.}
    \label{fig:ratios}
\end{figure}

We calculate the $\mathcal{D}_{\ell}^{\rm BB} / \mathcal{D}_{\ell}^{\rm EE}$, $\mathcal{D}_{\ell}^{\rm TE} / \mathcal{D}_{\ell}^{\rm EE}$, and $r_{\ell}^{\rm TE}$ ratios with our filament model and show them in Fig.~\ref{fig:ratios}. We show the ratios modeled by the Planck observation power law fits to each spectrum as dashed lines, as seen in Fig.~\ref{fig:ratios_planck_binning}, and we extend them to small scales, $\ell \sim 4000$. We can reproduce the tendency of power spectrum ratios measured at large scales by \citet{planck_2018_xi}, extrapolating them to higher multipoles.

We fit a power law to each ratio $R_{\ell}$, modeled by $R_{\ell} = A_R (\ell / 80 )^{\alpha_R}$ in the multipole range $1000 \leq \ell \leq 3000$. We use the Knox formula for the sample variance error bars and propagate them through the ratio division. The fitted amplitudes $A_R$ and slopes $\alpha_R$ are listed in Table~\ref{table:fitted_quantities}. The power laws fitted directly to the ratios are the solid lines in Fig.~\ref{fig:ratios}.

\begin{deluxetable*}{lllll}
    \tablecaption{Fitted quantities in our filament model. \label{table:fitted_quantities}}
    \tablehead{ \colhead{Quantity} & \colhead{Model} & \colhead{Multipole Range}	& \colhead{Fitted Amplitude} & \colhead{Fitted Slope} }
    \startdata
    \\
    \multicolumn{5}{c}{LR71 I,P Masks ($N_{\rm side}= 512$)} \\
    \hline
    $D_{\ell}^{\rm TT}$ & $A_{\rm TT}(\ell/80)^{\alpha_{\rm TT}+2}$ & $300\leq\ell<600$ \tablenotemark{a} & $A_{\rm TT}=28308\pm1443$\,$\mu$K$^2$ & $\alpha_{\rm TT}=-2.62\pm0.03$ \\
    \hline
    $D_{\ell}^{\rm EE}$ & $A_{\rm EE}(\ell/80)^{\alpha_{\rm EE}+2}$ & $40\leq\ell<600$ \tablenotemark{a} & $A_{\rm EE}=346\pm8$\,$\mu$K$^2$ & $\alpha_{\rm EE}=-2.50 \pm 0.02$ \\
    $D_{\ell}^{\rm BB}$ & $A_{\rm BB}(\ell/80)^{\alpha_{\rm BB}+2}$ & $40\leq\ell<600$ \tablenotemark{a} & $A_{\rm BB}=198\pm6$\,$\mu$K$^2$ & $\alpha_{\rm BB}=-2.65\pm0.02$ \\
    $D_{\ell}^{\rm TE}$ & $A_{\rm TE}(\ell/80)^{\alpha_{\rm TE}+2}$ & $40\leq\ell<600$ \tablenotemark{a} & $A_{\rm TE}=792\pm31$\,$\mu$K$^2$ & $\alpha_{\rm TE}=-2.48\pm0.02$ \\
    \hline
    \\
    \multicolumn{5}{c}{DR2 Galactic Mask with $f_{\rm sky}=0.7$  ($N_{\rm side}= 2048$)} \\
    \hline
    $D_{\ell}^{\rm TT}$ & $A_{\rm TT}(\ell/80)^{\alpha_{\rm TT}+2}$ & $1000\leq\ell<3000$ & $A_{\rm TT}=23923 \pm 246$\,$\mu$K$^2$ & $\alpha_{\rm TT}=-2.633\pm0.003$ \\
    $D_{\ell}^{\rm EE}$ & $A_{\rm EE}(\ell/80)^{\alpha_{\rm EE}+2}$ & $1000\leq\ell<3000$ & $A_{\rm EE}=281\pm4$\,$\mu$K$^2$ & $\alpha_{\rm EE}=-2.459\pm0.004$ \\
    $D_{\ell}^{\rm BB}$ & $A_{\rm BB}(\ell/80)^{\alpha_{\rm BB}+2}$ & $1000\leq\ell<3000$ & $A_{\rm BB}=152\pm2$\,$\mu$K$^2$ & $\alpha_{\rm BB}=-2.590\pm0.003$ \\
    $D_{\ell}^{\rm TE}$ & $A_{\rm TE}(\ell/80)^{\alpha_{\rm TE}+2}$ & $1000\leq\ell<3000$ & $A_{\rm TE}=727\pm16$\,$\mu$K$^2$ & $\alpha_{\rm TE}=-2.511\pm0.007$ \\
    \hline
    $D_{\ell}^{\rm BB}/D_{\ell}^{\rm EE}$ & $A_{BBEE}(\ell/80)^{\alpha_{BBEE}}$ & $1000\leq\ell<3000$ & $A_{BBEE}=0.543\pm0.008$ & $\alpha_{BBEE}=-0.131\pm0.005$ \\
    $D_{\ell}^{\rm TE}/D_{\ell}^{\rm EE}$ & $A_{TEEE}(\ell/80)^{\alpha_{TEEE}}$ & $1000\leq\ell<3000$ & $A_{TEEE}=2.59\pm0.05$ & $\alpha_{TEEE}=-0.053\pm0.006$ \\
    $r_{\ell}^{\rm TE}$ & $A_{r^{\rm TE}}(\ell/80)^{\alpha_{r^{\rm TE}}}$ & $1000\leq\ell<3000$ & $A_{r^{\rm TE}}=0.280\pm0.005$ & $\alpha_{r^{\rm TE}}=0.035\pm0.006$ \\
    \enddata
    \tablenotetext{a}{This binning scheme is listed in Table C.1 of \citet{planck_2018_xi}. 
    }
\end{deluxetable*}

\subsection{SED decorrelation} \label{sec:decorrelation_results}

\begin{figure}
    \centering
    \includegraphics[width=1.0\columnwidth]{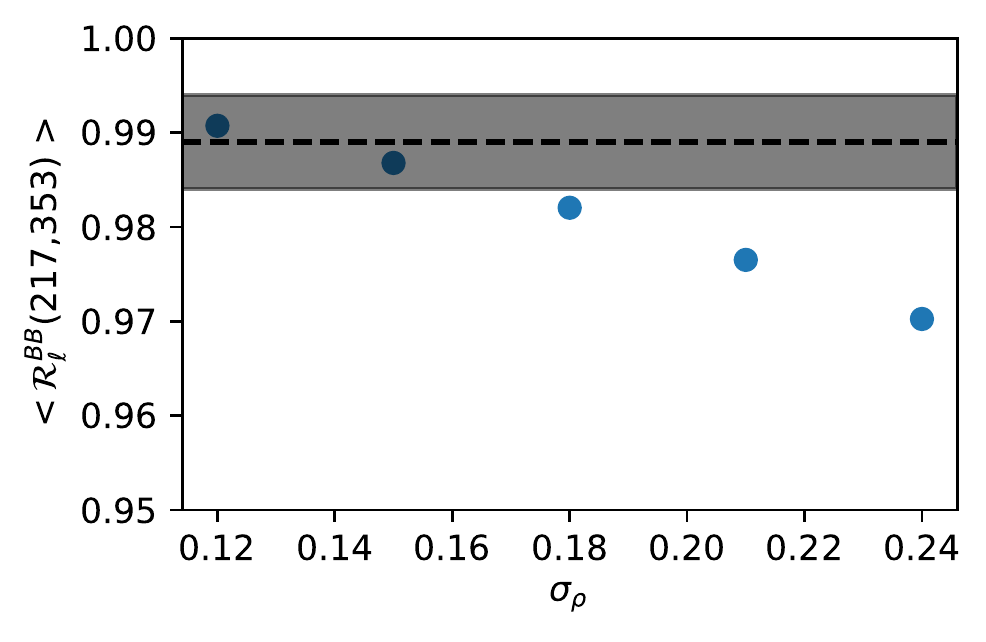}
    \caption{Frequency decorrelation ratio $\mathcal{R}_{\ell}^{\rm BB}(217,353)$, measured with the LR71 mask, versus the standard deviation $\sigma_{\rho}$ (eq.~\ref{eq:perturbation}) used to generate variability on the dust SED along an LOS. \citet{2021A&A...647A..16P} found that $\sigma_{\rho}=0.15$ works for reproducing decorrelation along individual LOSs. The $\mathcal{R}_{\ell}^{\rm BB}(217,353)$ at small scales from our model is scale-independent, so we show its mean value at small scales, calculated in the range $\ell=320-1500$. The dashed horizontal line is the value measured in \citet{planck_2018_xi} in the multipole range $50 \leq \ell < 150$, $\mathcal{R}_{\ell}^{\rm BB}(217,353)=0.989 \pm 0.005$.
    }
    \label{fig:sed_decorrelation}
\end{figure}

We produce a full-sky map of our filament model at 217\,GHz. We apply a distinct MBB spectral law to each individual filament, as explained in Section~\ref{sec:sed_method}. We generate a random realization for $\beta_{\rm dust}$ for each filament following eq.~\ref{eq:beta-dust}. The $\delta$ ratio between the 217 and 353\,GHz MBB and the $T_{\rm dust}$ parameters are set to the value at the pixel on which the center of each individual filament is located. The full-sky maps of $\delta$ and $T_{\rm dust}$ are calculated from the best-fit $\beta_{\rm dust}$ and $T_{\rm dust}$ maps calculated by the GNILC method in \citet{2016A&A...596A.109P}.

\citet{2021A&A...647A..16P} found that $\sigma_{\rho} = 0.15$ can reproduce the level of variability seen in the LOSs near the Galactic poles. We use that value, together with other values $\sigma_{\rho}=$0.12, 0.18, 0.21, and 0.24 to see the effect. We measure the degree of spectral decorrelation with the BB spectral correlation ratio $\mathcal{R}_{\ell}^{\rm BB}(217,353)$, defined in \citet{2017A&A...599A..51P} as

\begin{equation}
    \mathcal{R}_{\ell}^{\rm BB}(217,353) = \frac{\mathcal{D}_{\ell}^{\rm BB}(217\times353)}{\sqrt{\mathcal{D}_{\ell}^{\rm BB}(217\times217) \mathcal{D}_{\ell}^{\rm BB}(353\times353)}} \text{,}
\end{equation}
where 217 and 353 represent the 217 and 353\,GHz Planck frequency maps.

In Fig.~\ref{fig:sed_decorrelation} we show the mean $\mathcal{R}_{\ell}^{\rm BB}$ ratio calculated in our filament model with different values of $\sigma_{\rho}$ using mask LR71. Since this ratio is calculated in polarization, the large scale $\mathcal{R}_{\ell}^{\rm BB}$ ratio is influenced by the filling of the large-scale with the Planck template, as described in Section~\ref{sec:fill-large-scale}. 
All of the emission at $\ell < 50$ is completely due to the Planck template, while the emission in the range $50 \leq \ell < 300$ is a mixture of the Planck template and our filament model, depending on the ad hoc filter defined in eq.~\ref{eq:template-filter} and shown in Fig.~\ref{fig:ad hoc-filter}. The emission at $\ell > 300$ is completely due to our filament model. In these small scales, the decorrelation ratio from our filament model is scale-independent, so we calculate the mean $\mathcal{R}_{\ell}^{\rm BB}$ in the range $320 \leq \ell < 1500$ and plot that versus $\sigma_{\rho}$. We also include the measured ratio $\mathcal{R}_{\ell}^{\rm BB}=0.989\pm0.005$ by \citet{planck_2018_xi} in the mask LR71 and multipole range $50 \leq \ell < 150$. Here $\sigma_{\rho}=0.15$ seems to produce a realistic ratio $\mathcal{R}_{\ell}^{\rm BB} > 0.98$. By increasing $\sigma_{\rho}$, we increase the variability of the random $\beta_{\rm dust}$ along each LOS and increase the frequency decorrelation by lowering the $\mathcal{R}_{\ell}^{\rm BB}$ ratio almost linearly to any desired value.

As noted by \citet{2021A&A...647A..16P}, SED frequency decorrelation is not uniform throughout the sky but rather depends on the 3D structure of Galactic dust clouds and the Galactic magnetic field. In our filament model, we assume that filaments have distinct SEDs and therefore the LOS effect will create decorrelation, but spatially uniform since we assume the same $\sigma_{\rho}$ independent of how many filaments are located in a given LOS. This produces a scale-independent $\mathcal{R}_{\ell}^{\rm BB}$ ratio, as noted above. As such, our way of modeling frequency decorrelation is a crude approximation that can roughly reproduce the overall $\mathcal{R}^{\rm BB}_{\ell}$ ratio measured by Planck, but it cannot reproduce the individual LOS frequency decorrelation features.

\subsection{Density of filaments, Non-Gaussianity and MFs} \label{sec:MFs}

\begin{figure}
    \centering
    \includegraphics[width=1.0\columnwidth]{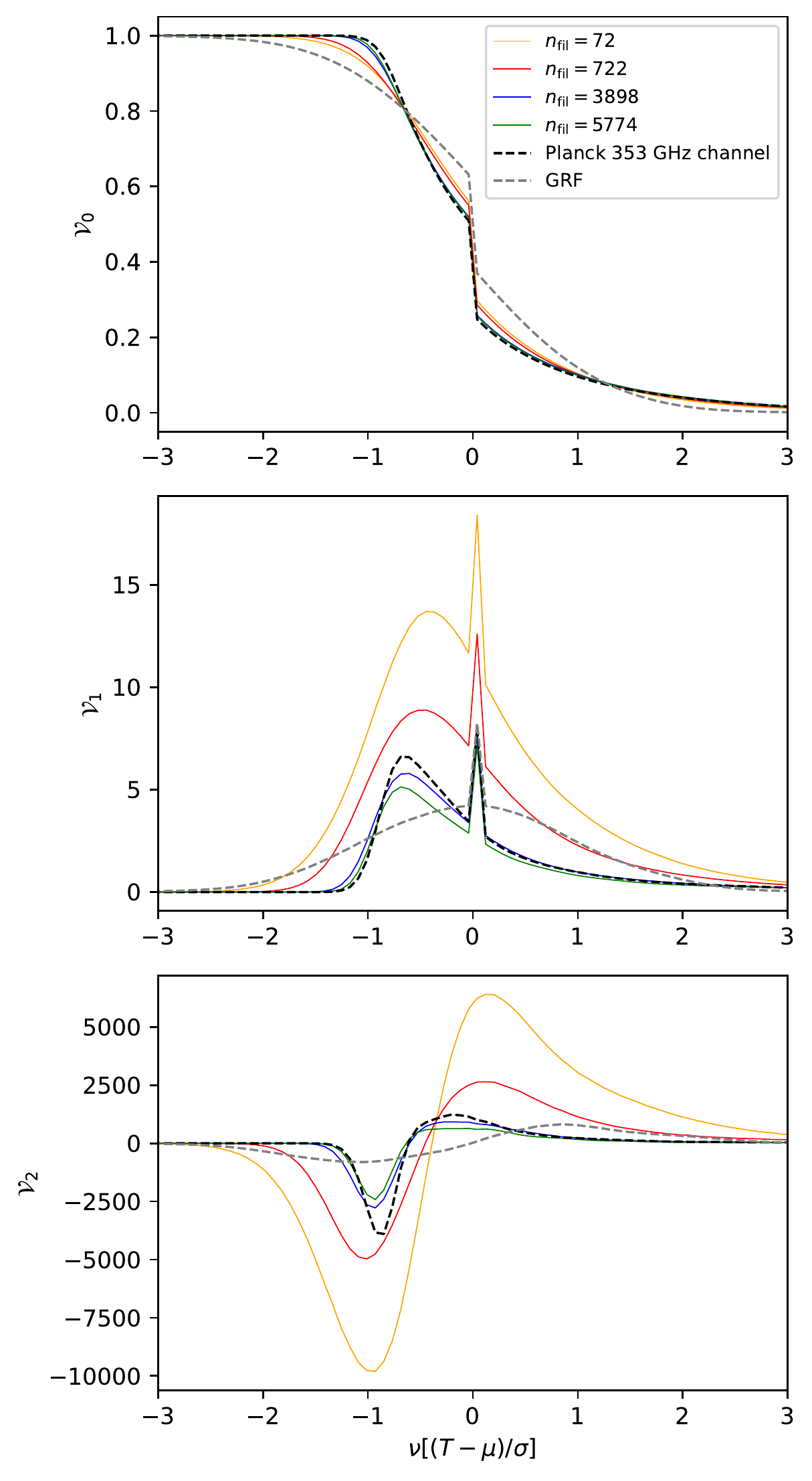}
    \caption{The MFs calculated over the $T$ map within the LR71 mask show a reasonable agreement between the filament model and Planck data. The dashed black line is the dust data: the Planck 353\,GHz frequency channel full mission map with the SMICA CMB map subtracted.  The dashed gray line is a GRF with the power law fit of the Planck $D_{\ell}^{\rm TT}$.  The solid colored lines are our filament model with different filament densities.  The blue line is the best fit we obtained by matching to the power spectrum, independent of the MFs.  The unit of $n_{\rm fil}$ is deg$^{-2}\times [I_{\rm dust}/({\rm MJy\ sr}^{-1})]$.  To make the comparison, we smooth our filament model and the GRF with the Planck 353\,GHz beam and add a noise simulation for that channel. The threshold $\nu$ is normalized by subtracting the mean and dividing by the $\sigma$ of the map.  The jumps at $\nu=0$ are due to the zero-value pixels in the mask.
    }
    \label{fig:MFs}
\end{figure}


Our filament model depends on the filament density $n_{\rm fil}$ of our population, which ultimately depends on the total number of filaments $N_{\rm fil}$ to achieve this.  A low density of filaments will render a highly non-Gaussian field, but due the central limit theorem, we expect that as $n_{\rm fil} \rightarrow \infty$, the field will get closer to Gaussian.


We wish to examine the relationship between Gaussianity and the $n_{\rm fil}$ parameter and compare to the Gaussianity in the Planck 353\,GHz map.   We focus on temperature because, in polarization, the observations have a low signal-to-noise ratio outside the Galactic plane at small scales; therefore, it is very hard to constrain the non-Gaussianity \citep[e.g.][]{2019MNRAS.487.5814V}.  Those sky areas are noise-dominated at the pixel level in polarization.

As explained in Section~\ref{sec:normalization}, the relative power of the one- and two-filament terms already determines $n_{\rm fil}$ by fitting to the Planck-measured $\mathcal{D}_{\ell}^{\rm TT}$ spectrum, and we found that the best-fit density is $n_{\rm fil} = 3898$\,deg$^{-2}\times [I_{\rm dust}/({\rm MJy\ sr}^{-1})] = n_{\rm fil}^{\rm best}$.

We use MFs \citep{1994A&A...288..697M} to quantify the level of non-Gaussianity and directly compare to the Planck 353\,GHz observations.  We use the 353\,GHz frequency channel map, since it has the highest resolution ($\sim 5\arcmin$) and signal-to-noise ratio for dust of the polarized Planck channels.  The three MFs, $\mathcal{V}_0(\nu)$, $\mathcal{V}_1(\nu)$, and $\mathcal{V}_2(\nu)$, measure the area, the perimeter length, and the genus of the excursion set at threshold $\nu$ in a map.
The genus equals the total number of connected regions above a given contour level $\nu$ minus the number of connected regions below.

We compute the three curved-sky MFs \citep[calculated via code from][]{2019JCAP...06..019M} within the LR71 mask for three kinds of maps: 
(1) the $T$ map for the Planck 353\,GHz full mission frequency channel \citep[with the best-fit SMICA CMB map subtracted;][]{2020A&A...641A...4P};
(2) a Gaussian random field (GRF) generated with the $\mathcal{D}_{\ell}^{\rm TT}$ power law fit of the Planck 353\,GHz map, as described in Section~\ref{sec:data}; and 
(3) our filament model with different values for the filament density $n_{\rm fil}$. Since the Planck 353\,GHz map has the instrument beam and noise in it, we have to apply the same to our filament model and the GRF.  We smooth the synthetic maps with the 353\,GHz channel beam in harmonic space, and then we add one of the 353\,GHz channel noise realizations from the Planck FFP simulations. All maps are normalized by subtracting the mean and dividing by the standard deviation.  The zero-valued pixels in the mask create a jump and spike at $\nu = 0$ for $\mathcal{V}_0$ and $\mathcal{V}_1$, respectively.

We also checked the consistency of the calculated MFs against the \textsc{cnd\_reg2d} code \citep{2013MNRAS.429.2104D}.

Figure~\ref{fig:MFs} shows that our filament model with $n_{\rm fil}=n_{\rm fil}^{\rm best}$ (blue) fits well the Planck 353\,GHz map (dashed black). We can see the highly non-Gaussian maps with a very low density, $n_{\rm fil}=72$ deg$^{-2}\times [I_{\rm dust}/({\rm MJy\ sr}^{-1})]$, and how by increasing the filament density, we approach the non-Gaussianity levels of the Planck 353\,GHz map. Also, both the Planck 353\,GHz map and our filament models are very distinct from a GRF.

We perhaps should not read too much into this agreement, since we are placing the filaments to mimic the large-scale features in the temperature map, and by design, the filament model reproduces its power spectrum.
As noted, adjustments to the filament density $n_{\rm fil}$ modify the power spectrum, which will modify the MFs by changing the overall variance, even if they had no other effects.


To focus on the scales more directly generated by the filaments, as a second test, we computed the MFs while limiting the range of scales with a bandpass harmonic filter that allows only $\ell=300-1200$,  where we have checked that the Planck 353\,GHz signal-to-noise ratio is $>1$. Our filament model is non-Gaussian on these scales compared to the GRF, but the Planck 353\,GHz map is substantially more non-Gaussian than our model on those scales.



\section{Discussion} \label{sec:discussion}

By design, our model reproduces well the power spectra of dust within the large-area LR71 Planck mask.  By mimicking the large-scale structure of the Planck dust intensity at the same time, it also reproduces the MF statistics on large scales.  However, the following effects cause differences between our models and the real sky that may be relevant for some applications.

\subsection{Large-scale polarization fraction fluctuations} The polarization fraction of thermal dust varies greatly across the sky (see, e.g., Fig.~43 of \citet{2016A&A...594A..10P} or Fig.~4 of \citet{2015A&A...576A.104P}), ranging from zero to $\sim 0.2$ in the $80\arcmin$ GNILC dust template, with a substantial uncertainty due to estimation of the dust monopole.  In our model, two mechanisms cause the polarization fraction to vary. First, the summation of polarization from large numbers of filaments along dense LOSs depolarizes the total signal.  This effect is seen near the Galactic plane, and will tend to make our model more polarized toward the Galactic poles, where there are fewer filaments.

Second, geometric effects cause filaments aligned to the LOS to be less polarized (section \ref{sec:pol-fraction}), so the particular realization of the large-scale magnetic field is also important.  The polarization fraction in maps produced by our model varies between zero and $\sim 0.2$\footnote{This range somewhat depends on the seed for the random magnetic field, since there could be particularly bad luck realizations that will render relatively highly polarized filaments.} at $80\arcmin$ resolution, and the geometry is different.  The polarization fraction of the raw filament map has more variability than the GNILC dust template when filtered similarly.  Our model also tends to have a higher polarization fraction toward the Galactic poles, as described above.

\subsection{Spatially varying physical polarization conditions} Table~1 of \citet{planck_2018_xi} shows that the $r_{\ell}^{\rm TE}$ and $\mathcal{D}_{\ell}^{\rm TE}/\mathcal{D}_{\ell}^{\rm EE}$ ratios change in masks with varying sky fractions.  We have argued that these quantities depend on the distribution of the intrinsic polarization fraction per filament, but in our model, we do not allow these distributions to vary as a function of position.  As a consequence, the power spectrum ratios that depend on the polarization change across the sky in the observations, but they stay constant in our model by construction. 


\subsection{Consequences for modeling small, clean patches}

When our model is calibrated to match the power spectrum from the overall LR71 sky region, that power is dominated by the brightest regions.  Thus calibrated, our model can have trouble reproducing the polarization power spectrum in smaller, cleaner regions.  This can be because of the above points about nonuniversal polarization behavior, as well as issues with the dust intensity template that underlies our distribution of filaments. 



The GNILC dust template has a higher specific intensity $\langle I_{353} \rangle$ than reported in Table 1 of \citet{planck_2018_xi} for the different sky fractions. To match this, you can subtract the dust monopole, but this is uncertain, since the largest scales are subjected to more systematics due to calibration drifts, etc. Subtracting a monopole has more impact in the cleanest regions of the sky, close to the Galactic poles. In these regions, the monopole represents a significant fraction of the emission, while regions with bright dust are little affected.


As a concrete example, in the BICEP/KECK (BK) region, we find that the EE and BB spectral amplitude of our model's default realization is about 10 times the amplitude observed in the sky. We note in particular that, in Planck data, the polarization fraction in the BK patch is lower than the average overall LR71 mask region.  Through geometric effects, the polarization fraction in our model depends on the random realization of the local magnetic field.  By chance in our default realization, the polarization fraction is higher than average in the BK patch, which contributes to our polarization power discrepancy there.

To match the power spectrum in a specific, small sky region, the raw maps from the filament model can be normalized to any spectra, as described in Section~\ref{sec:normalization}. If the objective is to use our model in the BK region, for example, the raw model can be normalized to temperature units with EE and BB spectra that match the observations in the BK region, therefore matching the observed amplitude of polarization while generating a filament realization. 

\section{Summary and conclusions} \label{sec:conclusions}

In this work, we present a model for the millimeter thermal dust Galactic emission based on the idea of \citet{filament_paper} of using filaments that are misaligned with respect to the local magnetic field to recreate the observed TEB power spectra and ratios as measured by \citet{planck_2018_xi}. We produce a 3D population of millions of filaments and integrate the emission along the LOS to produce a full-sky \textsc{healpix} map of the Stokes parameters.

We produce maps at the nominal frequencies of ground-based experiments like SO and CMB-S4: 20, 27, 39, 93, 145, 225, and 280\,GHz. We also make available the maps at the Planck channels of 217 and 353\,GHz. We make these maps publicly available to the community, along with the \textsc{DustFilaments} code to produce them, at \url{https://github.com/chervias/DustFilaments}\footnote{\url{https://zenodo.org/badge/latestdoi/382487350}}. Our 353\,GHz map can be used as a template for the dust emission and scaled to any other frequency using, for example, an MBB spectral law. Our maps at other frequencies include frequency decorrelation as described in Section~\ref{sec:sed_method}.

Our model can reproduce the power law shape of the intensity and polarization spectra, as well as the $\mathcal{D}_{\ell}^{\rm BB}/\mathcal{D}_{\ell}^{\rm EE}$, $\mathcal{D}_{\ell}^{\rm TE}/\mathcal{D}_{\ell}^{\rm EE}$, and $r_{\ell}^{\rm TE}$ ratios as observed by Planck. Our final thermal dust model is produced at $N_{\rm side}=2048$, with a filament density $n_{\rm fil}=3898$\,deg$^{-2}\times [I_{\rm dust}/({\rm MJy\ sr}^{-1})]$, which corresponds to $N_{\rm fil}=180.5$ million filaments for the full-sky. The box has a size $S=400$\,pc per side. The filament population is produced with $L_a^{\rm min}=0.01$\,pc, a Pareto distribution for the filament length $p(L_a) \propto L_a^{-2.445}$, field misalignment RMS$(\theta_{LH})=10$\degree, axis ratio $\epsilon(L_a) = 0.16 (L_a/L_a^{\rm min})^{+0.122}$, and a polarization fraction $f_{\rm pol} \propto f_{\rm pol,0} (L_a/L_a^{\rm min})^{-0.1}$, where $f_{\rm pol,0}$ is drawn from a beta distribution (with $\alpha=0.07734$ and $\beta=0.37448$ for the particular random realization of the magnetic field we use in this paper). We skip large filaments, using a size limit corresponding to $\ell_{\rm limit} = 50$. To generate the thermal dust SED with decorrelation, we allow the flux density ratio between the 217 GHz and 353 GHz bands to vary per filament with $\sigma_{\rho} = 0.15$. All of the parameters used to create our model are listed in Table~\ref{table:parameters}.

This modeling reproduces the spectra in the large-area LR71 mask  by design. As explained in Section~\ref{sec:discussion}, polarization properties are not spatially homogeneous; therefore, our model can produce a mismatch in subregions, like the BK patch. Our model can always be renormalized to match the spectra in those regions.  In the future, we would like to improve the modeling so that we can match the large areas and small, clean regions simultaneously.


Our filament model offers novel features, including the relatively fast production of small-scale emission up to $\ell \sim 4000$, or even smaller scales by adjusting the minimum size and total number of filaments,  limited only by the computing power. Our filament model takes $\lesssim 5$ CPU hr per million filaments.
Our model is designed to naturally produce non-Gaussian emission at all scales. We compare directly with the DR3 353\,GHz frequency channel from Planck \citep{2020A&A...641A...3P} using MFs, and we are able to reproduce the general level of non-Gaussianity in the LR71 mask, although in detail, the MFs are not the same when comparing small scales in a bandpass filter. By switching the random seed, our model can produce a Monte Carlo realization of the full-sky dust emission in an $\sim$hour timescale on a small cluster, which can then be used in CMB experimental forecasting.


Recent works have looked at the impact of foregrounds on the small-scale lensing reconstruction, in particular the effect of non-Gaussian foreground residuals \citep[e.g.][]{2020JCAP...06..030B,2021arXiv210201045B}. One feature of our model is its ability to fine-tune the non-Gaussianity by increasing or decreasing the density of filaments or changing the filament profile. Our model can help with forecasting the performance of CMB weak-lensing reconstruction methods in the presence of highly non-Gaussian dust emission.  This will be very useful for exploring future methods both for the reconstruction itself and for the delensing of primordial B-mode surveys.

Our model can also impact studies of frequency decorrelation. Forecasts suggest that this could have a sizable impact on parametric component separation methods trying to clean the B-mode CMB observations to measure $r$ \citep{2018ApJ...853..127H}. Methods that model the dust SED using moment expansion seem promising in dealing with the extra complexities that frequency decorrelation brings \citep{2017MNRAS.472.1195C,2021A&A...647A..52M,2021MNRAS.503.2478R,2021JCAP...05..047A}. Our filament model can create dust frequency decorrelation that can be adjusted to any desired level with the $\sigma_{\rho}$ parameter and help to evaluate the performance of foreground cleaning methods for primordial B-mode surveys.

Regarding the $\mathcal{D}_{\ell}^{TB}$ and $\mathcal{D}_{\ell}^{EB}$ spectra, our model can easily be made to produce a signal that violates parity. As pointed out by \citet{filament_paper}, a preference in the handedness of the filament long axis with respect to the projected local magnetic field would create parity violation and nonzero TB and EB spectra. Recently, \citet{2021arXiv210500120C} tested this possibility by analyzing the misalignment between filamentary structure in intensity and the projected magnetic field. We will leave the modeling of nonzero TB and EB spectra for future work.

While our filament model produces non-Gaussian small-scale emission, it is not the same as the small-scale emission in the Planck 353\,GHz map, as measured by comparing bandpass-filtered maps. This could be due to multiple factors, such as the lack of realism brought by the modeling of the interaction between the ISM and the magnetic field, something that is studied by MHD simulations;  the fact that a filament Gaussian profile 
is not realistic on small scales; or even that filaments cannot explain the full picture of Galactic dust, and assuming that all dust particles are part of the filamentary structure may lead us to the wrong conclusions.
Other areas of our filament model can certainly be improved, since we make several approximations. For example, we place filaments in the celestial sphere according to dust templates, but the third dimension (the radial distance to the filament) is still drawn from a random sample. For example, we could make the model more realistic by using a 3D distribution of the dust density, measured by dust extinction \citep[e.g.][]{2017A&A...598A.125R, 2019MNRAS.483.4277C,2019ApJ...887...93G,2019A&A...625A.135L}, to place the individual filaments and trace the density structure of Galactic dust.

While the main purpose of our filament model is not to constrain the physical conditions of the local ISM and its magnetic field, using information on those conditions would certainly increase the realism of our model and presumably the match to Planck data. The fact that we cannot reproduce the large-scale polarization of dust is a consequence of not having a realistic model of the magnetic field in the local ISM. We know the field is irregular and does not follow the large-scale Galactic field \citep[e.g.][]{1999A&A...346..955L}. Some works have tried to model the local magnetic field traced by dust polarization \citep[e.g.][]{2018A&A...611L...5A,2020A&A...636A..17P}, and we leave to future work using such models to try to produce full-sky polarization maps that do not require their large-scale emission to be filled by a template.  Also in the future, we will test local ISM models and their match to the large-scale dust emission as measured by Planck.

To make a unified foreground model over a broad range of frequencies, we need to incorporate synchrotron emission and its correlation to dust \citep{2015JCAP...12..020C}, which we leave for future work.
We will also study the effect of non-Gaussian foreground residuals for the reconstruction of the small-scale CMB weak-lensing potential in the future, and we hope our dust model will contribute to the analysis of the effects of foreground non-Gaussianity and frequency decorrelation in the forecasting for upcoming CMB experiments.

\begin{acknowledgements}
We thank Gabriela Marques for providing the code to compute the curved-sky Minkowski functionals and for checking the consistency between her code and the \textsc{cnd\_reg2d} code. We thank Aditya Rotti for helping us in the measurement of the power spectrum slopes of the Planck data. We also thank David Collins and Kye Stalpes for useful conversations about the turbulent nature of the ISM.  Arthur Kosowsky helped us to understand how to describe the power spectrum of the isotropic magnetic field component. We thank Susan Clark and Brandon Hensley for providing comments on a draft of this work. We acknowledge support from NASA Astrophysics Theory Program award NNX17AF87G and NSF Astronomy and Astrophysics Grant program awards 1815887 and 2009870. This work uses the \textsc{healpix} \citep{2005ApJ...622..759G} and \textsc{namaster} \citep{2019MNRAS.484.4127A} software.
\end{acknowledgements}

\bibliography{biblio.bib}


\appendix

\section{Details of the method} \label{sec:appendix-method}

\subsection{Magnetic field model} \label{sec:magnetic-field}
The full Galactic magnetic field we use in our model has a large-scale regular component.  We also developed code to add an isotropic random component.

\subsubsection{Large-scale Galactic component} \label{sec:large-scale-magnetic-field}
For the large-scale Galactic magnetic field, we use the model from \citet{2012ApJ...757...14J,2012ApJ...761L..11J}. This model has three components that describe the large-scale regular magnetic field in our galaxy, defined in galactocentric cylindrical coordinates.
\begin{itemize}
    \item \textit{Disk component.} The disk magnetic field only has components in the azimuthal and radial directions, so the Z component is null. This disk defines eight logarithmic spirals.
    \item \textit{Halo component.} This magnetic field only has an azimuthal component, which drops exponentially with the Z height up and down with respect to the Galactic plane disk.
    \item \textit{Out-of-plane component.} This component is also called the X halo, since the magnetic field lines resemble an X when observed from the edge of the Galactic disk. This has been observed in external galaxies. 
\end{itemize}

We adopt a cube with side $S=400$\,pc, larger than the $\sim 100$\,pc diameter Local Bubble and big enough to include some fraction of the Galactic neighborhood. We use the Jansson \& Farrar model shifted 8.5\,kpc along the X-axis to place the cube around the solar system, with the Galactic center toward the +X direction. Our $\bm{H}$ cube has a resolution of $256^3$ voxels.

\subsubsection{Generating the isotropic random magnetic field} \label{sec:code-magnetic-field}

Following \citet{2002PhRvD..65l3004M}, the power spectrum of a homogeneous, isotropic magnetic field is
\begin{equation}
  \langle H_i(\mathbf{k})  H_j(\mathbf{k}')^* \rangle = (2\pi)^3 {\cal P}_{ij}(\mathbf{k}) P(k) \delta(\mathbf{k - k}')
  \label{eqn:psdef}
 \end{equation}
 where $ij$ are the vector components of the field, $\mathbf{k}$ is the wavevector of the mode, and the projection operator to the transverse plane is
 \begin{equation}
    {\cal P}_{ij}(\mathbf{k})  = \delta_{ij} - \hat k_i \hat k_j.
 \end{equation}
 Here $\hat k_i$ are the components of the unit vector in the longitudinal ($\mathbf{k}$) direction.  To fill our simulation box with a magnetic field, we generate harmonics that satisfy the above relationship via
  \begin{equation}
  H_i (\mathbf{k}) = \left[ \frac{(2\pi)^3 P(k)}{\Delta_{{\rm vol}\,k}} \right]^{1/2} {\cal P}_{ij}(\mathbf{k}) g_j(\mathbf{k}),
\end{equation}
where $g_j$ is a vector of complex Gaussian random deviates with unit variance in each component, and $\Delta_{{\rm vol}\,k}$ is the volume of a pixel in harmonic space.  We transform the field components to real space with a fast Fourier transform.

With this model, we aim to simulate the isotropic random component seen in the small-scale Galactic magnetic field \citep{2015ASSL..407..483H}. We generate our random magnetic field  cube component with a power law spectrum $P(k) \propto k^{-4}$, which corresponds to roughly the small-scale spectrum seen in MHD simulations of the ISM (Stalpes et al. 2022 in preparation.). To add the large-scale Galactic component and the isotropic random component together, we normalize the random isotropic magnetic field cube so that RMS($|\bm{H}|$)$=3$\,$\mu$G \citep{2008A&A...477..573S,2010MNRAS.401.1013J}. 
 

\subsection{Defining the filament population} \label{sec:defining-filament-population}
The filaments are placed inside the cube defined by the model of the magnetic field. Because we do not want to generate distortion on the corners of this cube, we place the filaments inside a sphere centered at the observer with a radius equal to 0.45 times the side of the cube $S$. The positioning of the filaments inside this sphere can be accomplished in two main ways.

    \paragraph{Placement at random.} The centers of the filaments are generated randomly, such that the density of filaments per unit of volume is uniform. We generate random numbers $u \in \unif(-1,1)$, $\phi \in \unif(0,2\pi)$, and $r \in \unif(0.15,1)$. The random position of a filament in the Cartesian 3D space is then given by
    \begin{equation} \label{eq:placement}
        \bm{r} = R r^{1/3}(\sqrt{(1-u^2)\cos(\phi)} \bm{\hat{x}} + \sqrt{(1- u^2)\sin(\phi)} \bm{\hat{y}} + u \bm{\hat{z}}) \text{,}
    \end{equation}
    where $R$ is the radius of the sphere contained within the magnetic field cube. The random number $r$ is not generated starting at zero to avoid placing filaments that will overlap with the observer.
    \paragraph{Placement following a template map.} As described in Section~\ref{sec:properties-distribution-correlations}, for our primary method, we use a template map for placing the filament centers along the surface of the celestial sphere. This map can be a Galactic template or any other type, as long as it describes how intense each pixel is with respect to others. Each pixel will get a number of filaments given by a Poisson distribution with a parameter given by eq.~\ref{eq:lambda}. The third dimension, which is the radial distance to the filament, is randomly generated with $Rr^{1/3}$, just like it is done in equation~\ref{eq:placement}.  

\subsubsection{Orientation of the filaments} \label{sec:orientation-filaments}
The filaments are randomly oriented following the local magnetic field. The semi-major axis of the prolate spheroid is aligned with the local magnetic field at the center of the filament, $\bm{H}(\bm{r})$. The filament long axis is then rotated by a random angle $\theta_{LH}$ (with respect to some orthogonal vector) and then rotated again by a random angle $\phi \in \unif(0,2\pi)$ with respect to the local magnetic field $\bm{H}(\bm{r})$. The angle between the local magnetic field and the filament semi-major axis, $\theta_{\rm LH}$, is drawn from a random Gaussian distribution with zero mean and standard deviation RMS$(\theta_{LH})$. This way, the filaments will be approximately oriented with respect to the local magnetic field at their centers, and the degree of orientation is controlled by the RMS$(\theta_{LH})$ parameter. If $\bm{\hat{L}}$ is the unit vector along the semi-major axis of the filament, the orientation of the filament is described by two Euler angles, $\alpha_e$ and $\beta_e$ given by
\begin{align}
    \alpha_e &= \atantwo(\bm{\hat{L}}_y,\bm{\hat{L}}_x) \\
    \beta_e &= \arccos(\bm{\hat{L}}_z) \text{.}
\end{align}

\subsubsection{Sizes of the filament semi-major axis} \label{sec:sizes-filament-La}
The sizes of the filaments are generated randomly from a Pareto distribution, which is a power law $p(L_a) \propto L_a^{-\eta_L}$. As described in \citet{filament_paper}, this distribution will render the power law behavior observed by Planck in the various angular power spectra. The semi-major axis length $L_a$ is drawn from such a distribution, and the semi-minor axis length $L_b$ is defined as $L_b = \epsilon L_a$, where $\epsilon$ is the axis ratio defined for the population of filaments and the parameter that controls if the filaments are thick or thin. In principle, $\epsilon$ can be a constant number (i.e. all filaments have the same aspect ratio), dependent on the filament size, or stochastic. We also note that the central density, which could be proportional to the size, as described in Section~\ref{sec:los-integration}, will change the slope of the necessary Pareto distribution.
    
\subsubsection{Skipping large filaments} \label{sec:skipping-large-filaments}
One issue with the filament population that is not immediately obvious is the effect of very large filaments. The Pareto distribution for the filament major semiaxis length $L_a$ is a power law with a negative index, and as such, the longer side of the distribution does not have a hard bound, but rather the largest sample drawn will depend on the total number of filaments produced. When generating a population of several million filaments, some of those very large filaments will be generated, and they will be very prominent in the final Stokes parameter map. The largest filaments of all will be skipped when some fraction of their volume falls outside the modeled box, but other filaments will be slightly smaller than this threshold and will appear in the map. This feature is not realistic, so we do not add filaments to the final map whose projected angular size along the major axis $\Theta_a$ is larger than some predetermined scale.

The projected angular size $\Theta_a$ is given by
\begin{equation}
     \Theta_a = 2(L_a^2 \cos(\tan^{-1}(\frac{\tan(\theta_L)}{\epsilon}))^2 + L_b^2 \sin(\tan^{-1}(\frac{\tan(\theta_L)}{\epsilon}))^2)^{1/2}/r \text{,}
\end{equation}
where $\theta_L$ is the angle between the filament long axis and the LOS, and $r$ is the distance between the observer and the filament center. A filament has a smooth Gaussian profile, so it does not have a well defined edge. We choose to multiply by 2 since $L_a$ is the major semiaxis, and within two semiaxis length $2L_a$ in the Z-axis direction, a Gaussian profile for the filament density contains 68\,\% of the emission. We limit the scale to an appropriate value $\ell_{\rm limit}$, and we skip the filaments that fulfill the following condition
\begin{equation}
    \pi/\Theta_{a} < \ell_{\rm limit} \text{,}
\end{equation}
which effectively cuts off the one-halo term at low $\ell$.

\subsection{LOS integration} \label{sec:los-integration} 

Once we have a randomly generated filament population, we can integrate its emission along the LOS and project into the surface of the celestial sphere using the \textsc{healpix} conventions and tools. 

Every filament is defined within a rectangular box that extends to five times the size of the filament ($L_b$,$L_b$,$L_a$) in each axis direction ($X$,$Y$,$Z$). First, we need to define which LOSs will be integrated to paint an individual filament into the celestial sphere. Each LOS will correspond to a \textsc{healpix} pixel. We find which pixels are in the celestial sphere projection of an individual filament with the \emph{query polygon} method. Therefore, we need the coordinates of the outer perimeter of the projection of the rectangular box in the celestial sphere. To do this, we find the convex hull of the eight vertices that make up a filament box, projected in the 2D surface of the celestial sphere. We cannot use the latitude and longitude of each vertex straightforwardly as proxies for the X- and Y-coordinates, since we cannot control the behavior of filaments that cross either one of the poles or that cross the prime meridian. Instead, we transform the latitude-longitude coordinates into a stereographic projection, given by 
\begin{align}
    k &= 2/(1 + \sin(\pi/2 - \theta_c)\sin(\pi/2 - \theta_c) + \cos(\pi/2 - \theta_c)\cos(\pi/2 - \theta)\cos(\phi - \phi_c)) \\
    X &= k \cos(\pi/2-\theta)\sin(\phi - \phi_c) \\
    Y &= k [ \cos(\pi/2-\theta_c) \sin(\pi/2-\theta) - \sin(\pi/2-\theta_c) \cos(\pi/2 -\theta) \cos(\phi - \phi_c) ] \text{,}
\end{align}
where $\theta,\phi$ are the latitude and longitude of the corresponding vertex, and $\theta_c,\phi_c$ are the latitude and longitude of the center of the filament. With the convex hull, we determine which six of the eight vertices are on the exterior perimeter \footnote{Since we are calculating a convex hull with discrete floating point precision, we are prone to error when three or more points are very close to being colinear. In this case, we skip the particular filament, since the calculated convex hull might not be convex, and \emph{query polygon} will fail.}. We run the \emph{query polygon} function in this convex hull polygon to determine the list of pixels that belong inside the filament projection onto the celestial sphere.

For every LOS within each individual filament, we must determine at which distance the rectangular box is intersected going in and coming out. We do this by defining a set of coordinates $\bm{R}$ whose center is located at the center of the filament and rotated in conjunction with the orientation of the filament. To transform between the observer coordinates $\bm{r}$ and $\bm{R}$, we apply the rotation matrix $\bm{M}(\alpha_e,\beta_e)$ in the $Z_1Y_2Z_3$ convention,
\begin{equation} \label{eq:rotation-matrix}
    \bm{R} = \bm{M}(\alpha_e,\beta_e)\bm{r} + \bm{r}_{\rm c} = \left( \begin{smallmatrix} \cos(\alpha_e) \cos(\beta_e)&-\sin(\alpha_e)&\cos(\alpha_e)\sin(\beta_e) \\ \sin(\alpha_e)\cos(\beta_e)&\cos(\alpha_e)&\sin(\alpha_e)\sin(\beta_e) \\ -\sin(\beta_e)&0&\cos(\beta_e) \end{smallmatrix} \right) \bm{r} + \bm{r}_{\rm c} \text{,}
\end{equation}
where $\bm{r}_{\rm c}$ is the position of the center of the filament in the observer coordinates. At this stage, we determine if any of the eight vertices that define the filament rectangular box lie outside the magnetic field box, which could happen if the generated length $L_a$ is so long that the filament lies partly outside our defined box. In this case, we skip the filament, since the magnetic field is not defined inside it completely. We calculate the six normal unit vectors that are perpendicular to each of the six faces of the filament rectangular box, as well as the vectors that trace the four edges of each of the six faces. Now, we need to know the distance between the observer and one of the faces of the filament rectangular box across the LOS. To calculate this distance $r$, we intersect an LOS vector $\bm{\hat{r}}$ in the observer coordinates with each of the six faces of the filament rectangular box, using the equation for the intersection between a line and an infinite plane,
\begin{equation}
    r = \frac{\bm{p}_0 \cdot \bm{n}}{ \bm{\hat{r}} \cdot \bm{\hat{n}} } \text{,}
\end{equation}
where $\bm{p}_0$ is any point in the infinite plane defined by the face of the rectangular box (e.g. one of the vertices that belong to the particular face in question), and $\bm{\hat{n}}$ is the unit vector normal to the face (which is also normal to the infinite plane) in question. For every LOS, we do this six times, one for each face. We will then get six distances from the observer to the six infinite planes (in which the face is contained), but we only want two distances, one near and one far. If we define the four vectors $\bm{r}_{i}$ with $i \in {0,1,2,3}$ as the vectors pointing to the four corners of the face in question, we calculate the projection of the vector $r\bm{\hat{r}} - \bm{r}_{0}$ toward the two edge vectors $\bm{r}_{1} - \bm{r}_{0}$ and $\bm{r}_{3} - \bm{r}_{0}$. If the norm of each of the projections is above zero and below the norm of the respective vector $\bm{r}_{j} - \bm{r}_{0}$, with $j \in 1,3$, then we know that this particular face is intersected by the LOS vector. Finally, we know the two distances $r_{\rm near}$ and $r_{\rm far}$, one near and one far. Along the LOS vector, we know that between these two distances, the filament is defined. 

The next step is to integrate the LOS between $r_{\rm near}$ and $r_{\rm far}$. The Stokes parameters $T$, $Q$, and $U$ are defined by \citep[e.g.][]{2019ApJ...880..106K} 
\begin{align} \label{eq:TQU}
    T(\bm{\hat{r}}) &= A \int_{r_{\rm near}}^{r_{\rm far}} \rho_0 u(\bm{r})  dr \\
    Q(\bm{\hat{r}}) &= A f_{\rm pol,0} \sin^2\theta_H\int_{r_{\rm near}}^{r_{\rm far}} \rho_0 u(\bm{r}) \frac{\bm{H}_y(\bm{r})^2  - \bm{H}_x(\bm{r})^2 }{ |\bm{H}(\bm{r})|^2    } dr \\
    U(\bm{\hat{r}}) &= A f_{\rm pol,0} \sin^2\theta_H \int_{r_{\rm near}}^{r_{\rm far}} \rho_0 u(\bm{r}) \frac{-2 \bm{H}_x(\bm{r}) \bm{H}_y(\bm{r})}{|\bm{H}(\bm{r})|^2}  dr \text{,}
\end{align}
where $\bm{r} = r \bm{\hat{r}}$ is the radial vector along the LOS, and the $x,y$ subindex in the magnetic field $\bm{H}$ represents the projection along the two axes perpendicular to the LOS in the \textsc{healpix} convention (see Fig. 5 of the \textsc{healpix} primer \footnote{https://healpix.sourceforge.io/pdf/intro.pdf}). We choose the normalization $A$ to match the Planck power spectra (Section \ref{sec:normalization}).  We do not include the modulation in the dust intensity by the $\theta_H$ angle \citep{2019ApJ...887..159H}.
The density profile $u(\bm{r})$ is defined following the profile of the prolate spheroid. We define it in the $\bm{R}$ coordinates of the filament as a Gaussian profile,
\begin{equation}
    u(\bm{R}) = \exp(-\frac{1}{2}( (R_X/L_b)^2 + (R_Y/L_b)^2 + (R_Z/L_a)^2 )) \text{.}
\end{equation}
This can also be changed to any profile required. We transform between the $\bm{R}$ and $\bm{r}$ coordinates using the inverse of eq.~\ref{eq:rotation-matrix},
\begin{equation}
    \bm{r} = \bm{M}^{-1}(\alpha_e,\beta_e)(\bm{R} - \bm{r}_{\rm c}) \text{.}
\end{equation}
The normalization factor of the density, $\rho_0$, is set by Larson's law \citep{1981MNRAS.194..809L}, which states that the central density of star formation clouds is inversely proportional to the size of the cloud with a power law with index$\sim -1.1$. Therefore, we define 
\begin{equation}
    \rho_0 \propto L_a^{-1.1}
\end{equation}
for every filament. The local magnetic field, $\bm{H}(\bm{r})$, is interpolated using a trilinear interpolation, since the resolution of the magnetic field cube is limited. The polarization fraction $f_{\rm pol}$ is explained in Section~\ref{sec:pol-fraction}. We also include a computationally cheaper option to replace $\bm{H}(\bm{r})$ with $\bm{H}(\bm{r}_{\rm c})$, meaning that we do not interpolate the magnetic field at every position $\bm{r}$; instead, we calculate the magnetic field at the center of the filament $\bm{r}_{\rm c}$ only once per filament and apply that constant value throughout. 

\subsection{Variable sampling resolution} \label{sec:variable-sampling}

The wildly variable sizes of the filaments are a problem when we sample a \textsc{healpix} map with a fixed resolution. Filaments with a large angular size (either physically big or close to the observer) will be sampled with many pixels, often several million, while filaments with a very small angular size will be sampled with a handful of pixels, too few to accurately sample and average their elongated density profiles, which generates erroneous shot noise. To avoid these effects, we implement a variable resolution that keeps the pixel sampling nearly uniform relative to each filament. 

First, we choose an $N_{\rm side}^{\rm fixed}$ resolution parameter for the overall simulated map. Then, for every filament, we will determine a variable resolution parameter $N_{\rm side}^{\rm variable}$ and sample the LOS integration for that individual filament at that resolution. To determine this, we do the following calculation. Imagine we put the filament with its long axis aligned with the LOS. In this configuration, to the observer, the filament will look like a circle with a radius $L_b$, and the filament rectangular box will look like a square with a side $10 L_b$. We want to sample this side $10 L_b$ square with a grid of $N_{\rm reso}$ by $N_{\rm reso}$ pixels. In this case, the size of the pixel sampling this filament should be $\sim {(10 L_b)}/{(|\bm{\Vec{r}}_{\rm c}|N_{\rm reso})}$\,rad. The approximate size of a \textsc{healpix} pixel at resolution $N_{\rm side}$ is $\sqrt{{(4\pi)}/{(12N_{\rm side}^2)}}$. We want to choose $N_{\rm side}^{\rm variable}$ so that both of these pixel sizes are roughly equal. Then, the value of $N_{\rm side}^{\rm variable}$ is given by
\begin{equation}
    \log_2 N_{\rm side}^{\rm variable} = {\round{\log(0.1N_{\rm reso} \sqrt{\pi/3}|\bm{\Vec{r}}_{\rm c}|/L_b)/\log(2)}} \text{,}
\end{equation}
where $\round{}$ means round to the nearest integer, since $N_{\rm side}^{\rm variable}$ must be a power of 2. If $N_{\rm side}^{\rm variable} = N_{\rm side}^{\rm fixed}$, nothing special is required, and we add the filament map to the final map. 

If $N_{\rm side}^{\rm variable} > N_{\rm side}^{\rm fixed}$, then the filament map has a higher resolution than the fixed resolution. We need to down-sample the map. This is easily achieved thanks to the nested ordering in \textsc{healpix}, where every pixel at a higher resolution belongs to a parent pixel at a lower resolution. For every $M$ steps in resolution (for example, an $N_{\rm side}=2048-512$ change in resolution would be an $M=2$ step), every parent pixel at the lower resolution has $4^M$ children pixels at the higher resolution. Therefore, every pixel in the down-sampled map is the average of the $4^M$ children pixels from the high-resolution map. 

If $N_{\rm side}^{\rm variable} < N_{\rm side}^{\rm fixed}$, then we need to up-sample the map to the $N_{\rm side}^{\rm fixed}$ resolution.  Doing this up-sampling in pixel space is possible, but it is not recommended. The edges of the large parent pixels are visible in the higher-resolution map, and this creates undesirable small-scale artifacts in the power spectra. 
For this reason, we do this up-sampling in harmonic space using a Gaussian kernel with FWHM equal to the pixel size of an $N_{\rm side}^{\rm variable}$ map, which discards scales where such large filaments contribute little power and avoids the pixel effects.  Since harmonic space transforms are expensive, we add together all of the filaments sampled at the same resolution.  
At the end,  we up-sample these total filament maps to the fixed $N_{\rm side}^{\rm fixed}$ resolution, doing only one harmonic calculation per $N_{\rm side}$ between $128$ and $N_{\rm side}^{\rm fixed}/2$, and add them to the final map.

\section{E/B power from single filaments} \label{sec:single-filament-analysis} 

As noted by \citet{filament_paper}, the main parameters that control the relative E and B power in a filament are the misalignment angle between the filament long semiaxis and the magnetic field $\theta_{LH}$ and the axis ratio $\epsilon$. Changing the standard deviation of the misalignment angle RMS($\theta_{LH}$) will control how much correlation there is between the filaments and the magnetic field lines. A smaller RMS($\theta_{LH}$) means a higher degree of correlation, which generally means that the projected angle between the filament and magnetic field, $\psi_{LH}$, is smaller. This will decrease the $\mathcal{D}_{\ell}^{\rm BB}/\mathcal{D}_{\ell}^{\rm EE}$ ratio and increase the $r_{\ell}^{\rm TE}$ ratio. The latter will increase as the filament-magnetic field alignment increases, i.e. a smaller RMS($\theta_{LH}$). To understand the former effect, we look into the details of the spectra for a single filament.

\begin{figure}
    \centering
    \includegraphics[width=1.0\textwidth]{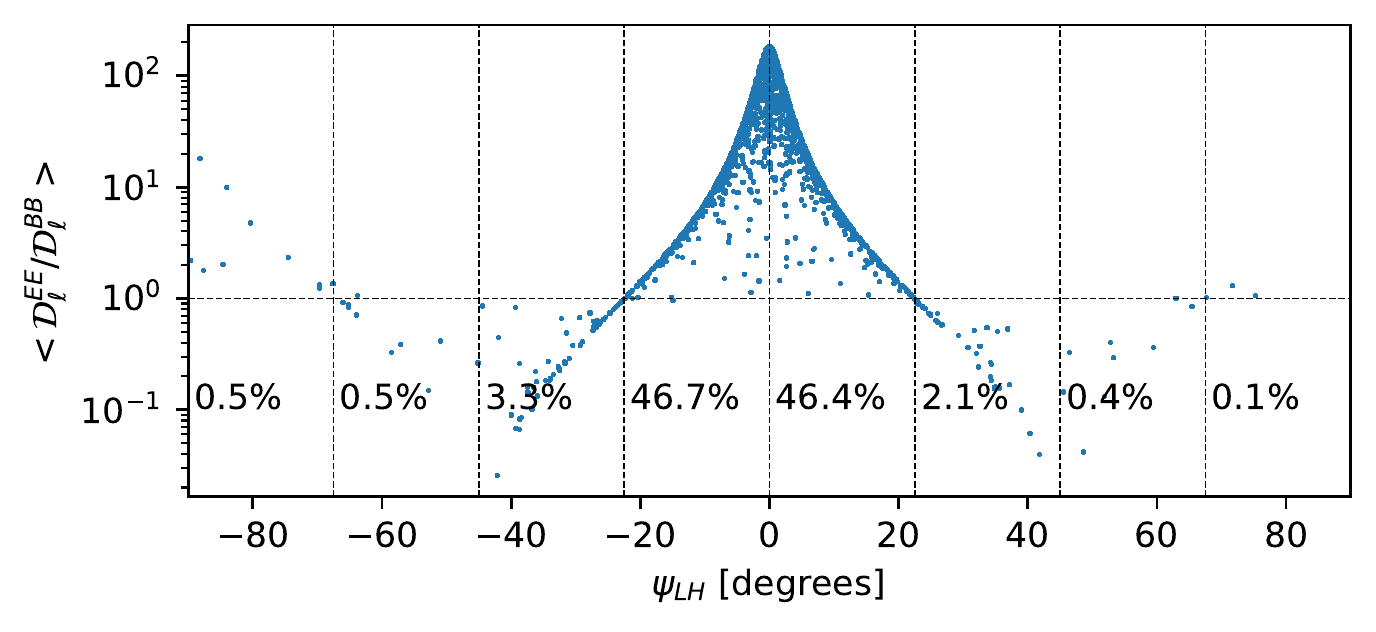}
    \caption{Mean small-scale $\mathcal{D}_{\ell}^{\rm EE}/\mathcal{D}_{\ell}^{\rm BB}$ ratio as a function of the projected angle between the filament and magnetic field $\psi_{LH}$ for a population of 2000 filaments, with rms($\theta_{LH}$)=10\degree, $\epsilon=0.16$, and filaments with the same length. The fraction of filaments that belong to each 22.5\degree bin in $\psi_{LH}$ is shown in each bin.}
    \label{fig:single-filament}
\end{figure}

Our starting point is Figure 2 of \citet{filament_paper}. We note that when the filament and magnetic field are aligned, the projected angle between them is $\psi_{LH}=0$.  There the E field reaches its maximum, and the B field reaches its minimum.  When there is a $\psi_{LH}=22.5$\degree angle between the filament and magnetic field, both the E and B fields have about the same power. Then, when the angle is $\psi_{LH}=45$\degree, E and B reverse roles: B reaches its maximum, and E reaches its minimum. This oscillation continues with a period of $90$\degree.

Fig.~\ref{fig:single-filament} shows this oscillation of the $\mathcal{D}_{\ell}^{\rm BB}/\mathcal{D}_{\ell}^{\rm EE}$ ratio as a function of the projected filament-magnetic field angle $\psi_{LH}$. We calculate the mean ratio for the small-scale spectra for an individual filament map. We use RMS($\theta_{LH}$)=10\degree and $\epsilon=0.16$ for 2000 filaments with equal length $L_a$. We note that for every 22.5\degree in $\psi_{LH}$, the ratio goes from minimum to maximum, or vice versa. Therefore, the key to achieving a filament population where the E-modes dominate over the B ones is to have more filaments in the $\psi_{LH}$ angle ranges where the ratio is above 1. We know that for $ | \psi_{LH} | < 22.5$\degree, the E-modes dominate, so filament populations with smaller rms($\theta_{LH}$) will have proportionally more filaments closer to alignment with the magnetic field and more E-mode domination. In the figure, the number on each $\psi_{LH}$ range shows the percentage of the 2000 filaments that belong to that range.

The dependence on the axis ratio $\epsilon$ is shown in Fig 1 of \citet{filament_paper}, where decreasing $\epsilon$, i.e. thinner filaments, increases the E-modes over the B-modes. The combined effect of fine-tuning  RMS($\theta_{LH}$) and $\epsilon$ in a filament population will render a given combination of $\mathcal{D}_{\ell}^{\rm BB}/\mathcal{D}_{\ell}^{\rm EE}$ and $r_{\ell}^{\rm TE}$ ratios.  The $r^{\rm TE}$ correlation can be further refined by tuning the distribution of the polarization fraction.

\end{document}